%% file: main.tex
\PassOptionsToPackage{dvipsnames,svgnames,table}{xcolor}
\documentclass[sigconf,nonacm,balance=false]{acmart}

\AtBeginDocument{%
  }

\input{config/packages.tex}

\input{config/math.tex}

\input{config/acronyms.tex}

\let\origparagraph\paragraph
\renewcommand{\paragraph}[1]{%
  \noindent\textbf{#1.}%
}
\usepackage{xcolor}
\definecolor{darkgreen}{RGB}{0, 100, 0}

\newcommand{\ours}[0]{{\sc Stretto}}

\begin{document}
\title{The \ours{} Execution Engine for Semantic Data Systems}

\author{Gabriele Sanmartino}
\authornote{These authors contributed equally to this work.}
\affiliation{%
  \institution{EURECOM}
  \city{Biot}
  \country{France}
}
\email{gabriele.sanmartino@eurecom.fr}

\author{Matthias Urban}
\authornotemark[1]
\affiliation{%
  \institution{Technical University of Darmstadt}
  \city{Darmstadt}
  \state{Hesse}
  \country{Germany}
}
\email{matthias.urban@tu-darmstadt.de}

\author{Paolo Papotti}
\affiliation{%
  \institution{EURECOM}
  \city{Biot}
  \country{France}
}
\email{papotti@eurecom.fr}

\author{Carsten Binnig}
\affiliation{%
  \institution{TU Darmstadt, DFKI, HessianAI}
  \city{Darmstadt}
  \state{Hesse}
  \country{Germany}
}
\email{carsten.binnig@tu-darmstadt.de}

\begin{abstract}
Semantic data systems enable querying over both structured and unstructured data using LLMs, but a central challenge is navigating the performance--accuracy trade-off when selecting among alternative strategies for executing semantic operators. In this paper, we present \ours{}, a new execution engine for semantic query processing that combines a novel optimizer with a rich set of physical execution strategies for semantic operators. Unlike prior work, \ours{} provides global --- rather than per-operator --- accuracy guarantees while optimizing query performance. To this end, \ours{} formulates query planning as a constrained optimization problem and employs a gradient-based optimizer to jointly select operator implementations and allocate error budgets across the full query plan.
Furthermore, to enable diverse execution choices, \ours{} introduces a new technique that builds on KV-cache compression to realize a spectrum of physical execution strategies for semantic operators with controllable runtime--accuracy trade-offs.
 Experimental results demonstrate that \ours{} outperforms state-of-the-art systems while consistently satisfying quality guarantees.\footnote{
Code and Artifacts  are available \href{https://github.com/DataManagementLab/stretto}{[here]}
}
\end{abstract}

\maketitle

\input{sections/introduction.tex}
\input{sections/overview}

\input{sections/optimization.tex}
\input{sections/operators.tex}

\input{sections/details}

\input{sections/experiments.tex}

\input{sections/related.tex}

\input{sections/conclusion.tex}

\begin{acks}
This work was funded by the DFG/ANR Project MAgiQ (ANR-24-CE92-0077; DFG, German Research Foundation – Project No. 545611510), the LOEWE Spitzenprofessur programme (III~5-519/\allowbreak 05.00.003-(0005)), by the Deutsche Forschungsgemeinschaft (DFG, German Research Foundation) under Germany’s Excellence Strategy (EXC-3057/1 “Reasonable Artificial Intelligence”, Project No. 533677015), and by the French government, through the 3IA Côte d’Azur Investments in the IA-cluster project managed by the National Research Agency (ANR-23-IACL-0001). We also thank DFKI and HessianAI.
\end{acks}

\bibliographystyle{ACM-Reference-Format}
\bibliography{biblio}

\newpage
\appendix

\clearpage
\input{sections/appendix.tex}
\let\paragraph\origparagraph

\end{document}

%% file: config/packages.tex
\usepackage{scalefnt,letltxmacro,tabularx}
\usepackage[acronym,nowarn]{glossaries}
\glsdisablehyper
\makeglossaries
\usepackage{xspace}
\usepackage{float}
\usepackage{physics}
\usepackage[normalem]{ulem}
\usepackage[english]{babel}

\usepackage{enumitem} %
\usepackage{booktabs} %
\usepackage{multirow} %
\usepackage{array}    %
\usepackage{wrapfig}  %
\usepackage{lipsum}   %
\usepackage[]{microtype} %
\usepackage{fancyvrb}
\usepackage{tabularx}
\usepackage{tablefootnote}
\usepackage{makecell} 
\usepackage{adjustbox}

\usepackage{amssymb}
\usepackage{amsmath}
\usepackage{amsfonts}
\usepackage{mathtools}

\usepackage{algorithm}
\usepackage{algpseudocode}

\usepackage[dvipsnames,svgnames,table]{xcolor}
\definecolor{pink3}{HTML}{fc0557}
\definecolor{shadecolor}{gray}{0.9}
\definecolor{mylightgray}{gray}{0.94}
\usepackage[all]{hypcap}
\hypersetup{
  citecolor=pink3,
  linkcolor=pink3,
  urlcolor=pink3
}
\usepackage[most,skins,theorems]{tcolorbox}

\usepackage[capitalize,nameinlink]{cleveref}
\crefname{section}{\S}{\S\S}
\Crefname{section}{\S}{\S\S}

\usepackage{graphicx}
\usepackage{subcaption}
\usepackage{tikz}
\usetikzlibrary{shapes.geometric}

\usepackage{pgfplots}
\pgfplotsset{compat=1.6}
\usepgfplotslibrary{groupplots}
\pgfplotsset{
  tick label style = {font=\sffamily},
  every axis label/.append style={font=\sffamily},
  typeset ticklabels with strut,
  xticklabel={$\mathsf{\pgfmathprintnumber{\tick}}$},
  yticklabel={$\mathsf{\pgfmathprintnumber{\tick}}$},
  every axis title/.append style={yshift=-1ex}
}

\usepackage[colorinlistoftodos,
  textsize=scriptsize,
  linecolor=BurntOrange,
  bordercolor=BurntOrange,
  backgroundcolor=BurntOrange!25]{todonotes}

\usepackage{multirow}
\usepackage{pifont}

\usepackage{url}

\tcbset{
  promptstyle/.style={
      enhanced,
      colback=white,
      colframe=black,
      colbacktitle=gray!20,
      width=0.98\linewidth,
      coltitle=black,
      rounded corners,
      sharp corners=north,
      boxrule=0.5pt,
      drop shadow=black!50!white,
      attach boxed title to top left={
          xshift=-2mm,
          yshift=-2mm
        },
      title code={%
          \begin{minipage}{0.8\linewidth}
            \raggedright\thetitle
          \end{minipage}
        },
      boxed title style={
          rounded corners,
          size=small,
          colback=gray!20,
        },
      fonttitle=\normalsize\bfseries,
      before upper={\parindent15pt\raggedright}
    }
}

\tcbset{
  aibox/.style={
      width=\linewidth,
      top=8pt,
      bottom=4pt,
      colback=blue!6!white,
      colframe=black,
      colbacktitle=black,
      enhanced,
      center,
      attach boxed title to top left={yshift=-0.1in,xshift=0.15in},
      boxed title style={boxrule=0pt,colframe=white,},
    }
}
\newtcolorbox{AIbox}[2][]{aibox,title=#2,#1}

\usepackage{listings}
\lstdefinestyle{TinyJSON}{
morekeywords={
    _question_, _evidence_, _database_schema_, im_start, im_end, user, system, assistant, <answer>, think, Wait,
  },
basicstyle=\ttfamily\footnotesize, %
keywordstyle=\color{purple}\bfseries, %
stringstyle=\color{purple}, %
commentstyle=\color{gray}\itshape, %
numbers=none, %
numberstyle=\tiny\color{gray}, %
stepnumber=1, %
numbersep=10pt, %
tabsize=2, %
captionpos=b, %
breaklines=true, %
breakatwhitespace=true, %
showspaces=false, %
showstringspaces=false, %
escapeinside={(*@}{@*)}, %
frame=none, %
extendedchars=true,
literate={á}{{\'a}}1 {ã}{{\~a}}1 {é}{{\'e}}1 {í}{{\'i}}1 {ó}{{\'o}}1 {ú}{{\'u}}1 {ç}{{\c{c}}}1,
}

%% file: config/math.tex
\DeclareMathOperator*{\argmin}{arg\,min}

%% file: config/acronyms.tex
\newacronym{SGD}{SGD}{stochastic gradient descent}
\newacronym{SGA}{SGA}{stochastic gradient ascent}
\newacronym{MAP}{MAP}{maximum-a-posteriori}
\newacronym{MLE}{MLE}{maximum likelihood estimation}
\newacronym{MNLL}{MNLL}{mean negative log-likelihood}
\newacronym{NLL}{NLL}{negative log-likelihood}
\newacronym{LL}{LL}{log-likelihood}
\newacronym{RMSE}{RMSE}{root mean square error}
\newacronym{ECE}{ECE}{expected calibration error}
\newacronym{SNR}{SNR}{signal-to-noise ratio}
\newacronym{FID}{FID}{Fr\'echet Inception Distance}
\newacronym{BPD}{BPD}{bit per dimension}
\newacronym{NFE}{NFE}{neural function evaluations}

\newacronym{AE}{AE}{autoencoder}
\newacronym{WAE}{WAE}{Wasserstein Autoencoder}
\newacronym{VAE}{VAE}{Variational Autoencoder}
\newacronym{BAE}{BAE}{Bayesian autoencoder}
\newacronym{GAN}{GAN}{Generative Adversarial Network}
\newacronym{DPGMM}{DPGMM}{Dirichlet process Gaussian mixture model}

\newacronym{MC}{MC}{Monte Carlo}
\newacronym{MCMC}{MCMC}{Markov chain Monte Carlo}
\newacronym{HMC}{HMC}{Hamiltonian Monte Carlo}
\newacronym{MH}{MH}{Metropolis-Hastings}
\newacronym{NUTS}{NUTS}{no-u-turn sampler}
\newacronym{SGHMC}{SGHMC}{stochastic gradient Hamiltonian Monte Carlo}
\newacronym{OU}{OU}{Ornstein-Uhlenbeck}

\newacronym{SDE}{SDE}{Stochastic Differential Equation}
\newacronym{CNF}{CNF}{Continuous Normalizing Flow}
\newacronym{ODE}{ODE}{Ordinary Differential Equation}
\newacronym{NF}{NF}{normalizing flow}

\newacronym{GP}{GP}{Gaussian Process}
\newacronym[longplural=deep Gaussian processes]{DGP}{DGP}{deep Gaussian process}
\newacronym{GPLVM}{GPLVM}{Gaussian process latent variable model}
\newacronym{DPMM}{DPMM}{Dirichlet Process Mixture Model}

\newacronym{LLM}{LLM}{large language model}
\newacronym{VI}{VI}{variational inference}
\newacronym{SVI}{SVI}{stochastic variational inference}
\newacronym{BNN}{BNN}{Bayesian neural network}
\newacronym{DNN}{DNN}{deep neural network}
\newacronym{CNN}{CNN}{convolutional neural network}
\newacronym{MLP}{MLP}{multilayer perceptron}
\newacronym{NN}{NN}{neural network}
\newacronym{RELU}{ReLU}{rectified linear unit}

\newacronym{ELBO}{ELBO}{evidence lower bound}
\newacronym{NELBO}{NELBO}{negative evidence lower bound}
\newacronym{ELL}{ELL}{expected log likelihood}
\newacronym{KL}{KL}{Kullback-Leibler divergence}
\newacronym{AUC}{AUC}{area under the curve}
\newacronym{VFE}{VFE}{variational free energy}

\newacronym{RBF}{RBF}{radial basis function}
\newacronym{ARD}{ARD}{automatic relevance determination}
\newacronym{RKHS}{RKHS}{reproducing kernel Hilbert space}

\newacronym{OT}{OT}{optimal transport}
\newacronym{WD}{WD}{Wasserstein distance}
\newacronym{SWD}{SWD}{sliced-Wasserstein distance}
\newacronym{DSWD}{DSWD}{distributional sliced-Wasserstein distance}

\newacronym{LAP}{LAP}{linear assignment problem}
\newacronym{SOLAP}{SOLAP}{sum of bilinear assignment problems}

\newacronym{ICL}{ICL}{in-context learning}
\newacronym{LoRA}{LoRA}{Low-Rank Adaptation}
\newacronym{SFT}{SFT}{Supervised Fine-Tuning}
\newacronym{RL}{RL}{reinforcement learning}
\newacronym{RLHF}{RLHF}{reinforcement learning from human feedback}
\newacronym{PPO}{PPO}{Proximal Policy Optimization}
\newacronym{GRPO}{GRPO}{Grouped Proximal Policy Optimization}
\newacronym{COT}{CoT}{chain-of-thought}

%% file: sections/introduction.tex
\section{Introduction}
\label{sec:intro}

\noindent \textbf{Semantic Data Systems enable Multi-Modal Querying.}
Semantic data systems integrate LLMs as first-class components of the data management stack, enabling semantic querying, extraction, and integration across both structured and unstructured data, including text, images, and audio. By integrating LLMs into query execution, these systems can execute queries directly over heterogeneous modalities.
For example, in a medical setting, a user may extract patient diagnoses from clinical reports and subsequently join them with structured patient metadata in a table. By eliminating the need to first transform unstructured data in reports into relational tables, this paradigm significantly lowers the barrier to querying complex data and is therefore highly attractive. %
Semantic data systems have recently gained substantial attention, with several systems %
demonstrating this approach~\cite{palimpzestCIDR,lotusVLDB,docEtlVLDB,thalamusdb,UrbanB24}.

\noindent \textbf{Execution of Semantic Queries.}
Executing semantic queries requires query engines capable of efficiently executing query plans that include \emph{semantic operators}, such as LLM-based filters, maps, and joins, alongside classical database operators. 
Unlike classical database operators, semantic operators expose a fundamentally different implementation design space, encompassing choices such as model sizes (e.g., smaller versus larger LLMs) and threshold settings for other implementation strategies, such as embedding similarity.
As with classical operators, the choice of physical implementation impacts the runtime of semantic operators. In addition, physical choices also influence result quality, introducing a runtime--accuracy trade-off that such systems must navigate.  

\noindent \textbf{The Missing Pieces.}
While semantic data systems have started to explore how to realize query engines with semantic operators, we argue that a more principled approach is required, one that enables these systems to select execution strategies that make efficient and systematic use of the runtime--accuracy trade-off. Overall, for finding optimal execution strategies, we argue that existing semantic data systems lack two major building blocks:

\noindent \textit{(1) No Global Quality Guarantees.}
Semantic data systems that support semantic operators have proposed initial techniques for selecting execution strategies that exploit the runtime--accuracy trade-off. 
For example, Lotus~\cite{lotusVLDB} was among the first systems to explicitly leverage this trade-off; however, it does so only \emph{locally}, optimizing each operator in isolation and providing quality guarantees only per operator. 
Such local guarantees can lead to suboptimal decisions, as they ignore how errors propagate across a pipeline of semantic operators.
Moreover, users ultimately care about the quality of the final query result rather than the accuracy of intermediate operators.
Other approaches, such as Abacus~\cite{abacusArxiv}, consider optimizations at the query level, but rely on heuristics to explore the plan space and also cannot provide guarantees on result quality.

In this paper, we argue for a global (holistic) approach to query optimization that reasons about accuracy end-to-end while still providing explicit guarantees to users.
Consider a medical researcher retrieving patients who exhibit both (a) \textit{abnormal lung scans} and (b) a textual diagnosis of \textit{COVID-19}, with a required global recall of $0.9$.
A naive strategy that splits this target symmetrically would require each filter to achieve a recall of $\sqrt{0.9} \approx 0.95$.
This requirement is expensive and potentially wasteful.
If filtering lung scans is comparatively easy (for instance, if a small vision-language model achieves near-perfect recall), then the text-based filter only needs to satisfy the global target of $0.9$ recall.
By failing to dynamically allocate the error budget across operators, a local optimizer may unnecessarily select a larger and more expensive LLM for the filter.

\noindent \textit{(2) No Principled Physical Operator Layer.}
Semantic operators in existing systems have been implemented using a variety of execution strategies.
Consider a semantic filter: systems %
employ different LLMs to implement it \cite{palimpzestCIDR}, %
combine a smaller LLM with a larger LLM in a so-called operator cascade \cite{lotusVLDB}, %
and employ different prompting strategies \cite{abacusArxiv, docEtlVLDB} or embedding-based similarity filters \cite{UrbanB24}. %
While these techniques expose a surface with different latency and accuracy trade-offs, the resulting search space of execution strategies for semantic operators is often noisy as it is unclear how the designs affect performance and accuracy, making it difficult for an optimizer to reason about the cost--accuracy trade-off in a more principled way.
That not only complicates query optimization due to the lack of a well-behaved search space, but also leads to coarse-grained gaps between available alternatives of executing semantic operators. As a result, an optimizer may be forced to select a significantly larger and therefore slower model than is strictly necessary to achieve a given accuracy target for a query.

\noindent \textbf{Our Approach: The \ours{} Execution Engine.}
In this paper, we present \ours{}, a new execution engine for semantic data systems that addresses the challenges outlined above. 
\ours{} consists of two complementary components: a novel global query optimizer and a novel physical operator layer for semantic operators.

First, unlike prior systems that rely on heuristics or local accuracy bounds, our optimizer in \ours{} explicitly provides global guarantees for a semantic query plan.
For this, it formulates query planning as a constrained optimization problem that minimizes global execution cost subject to end-to-end precision and recall constraints.
We solve this problem efficiently via a gradient-based search over a continuous relaxation of the plan space.
This formulation enables a unified exploration of the full plan space, jointly optimizing thresholds and model selections across all parameterized operators.
Crucially, it allows the optimizer to dynamically reallocate error budgets from ``easy'' operators to ``hard'' ones, thereby minimizing total cost while satisfying global quality guarantees.

Second, \ours{} introduces a new physical operator layer for semantic operators that enables principled navigation of the cost--accuracy trade-off.
Inspired by recent work on efficient LLM inference~\cite{pope2023efficiently, expectedattention}, we expose the LLM key--value (KV) cache as a first-class resource for semantic operators.
This allows us to construct a focused, controllable set of physical implementations that systematically trade runtime for accuracy.
Together, these two techniques (global optimization and a new physical operator layer) transform the large and noisy physical design space into a well-structured trade-off space.
This lets \ours{} dial in previously inaccessible trade-offs and thus provide better execution strategies, all while maintaining a streamlined and manageable operator space.

\noindent \textbf{Contributions.}
Overall, in this paper, we make the following contributions:
\emph{(1) Global gradient-based query optimization:} We propose a query optimizer that enforces end-to-end precision and recall constraints by formulating semantic query planning as a constrained optimization problem. It uses gradient-based search to jointly optimize operator selections and parameters while dynamically allocating error budgets across the pipeline.
\emph{(2) KV cache--based physical operator layer:} We introduce a new physical design for semantic operators that leverages compressed key--value (KV) caches to expose a smooth and fine-grained cost--accuracy spectrum, enabling more principled execution-time trade-offs.
\emph{(3) Systematic evaluation:} We implement \ours{} and show on diverse workloads, including SemBench~\cite{sembench}, that it achieves an average speed-up of $1.6\times$ over state-of-the-art baselines while satisfying global quality constraints.

%% file: sections/overview.tex
\vspace{-1ex}
\section{\ours{} Overview}
\label{sec:overview}

\begin{figure*}[h]
  \centering
  \includegraphics[width=0.8\linewidth]{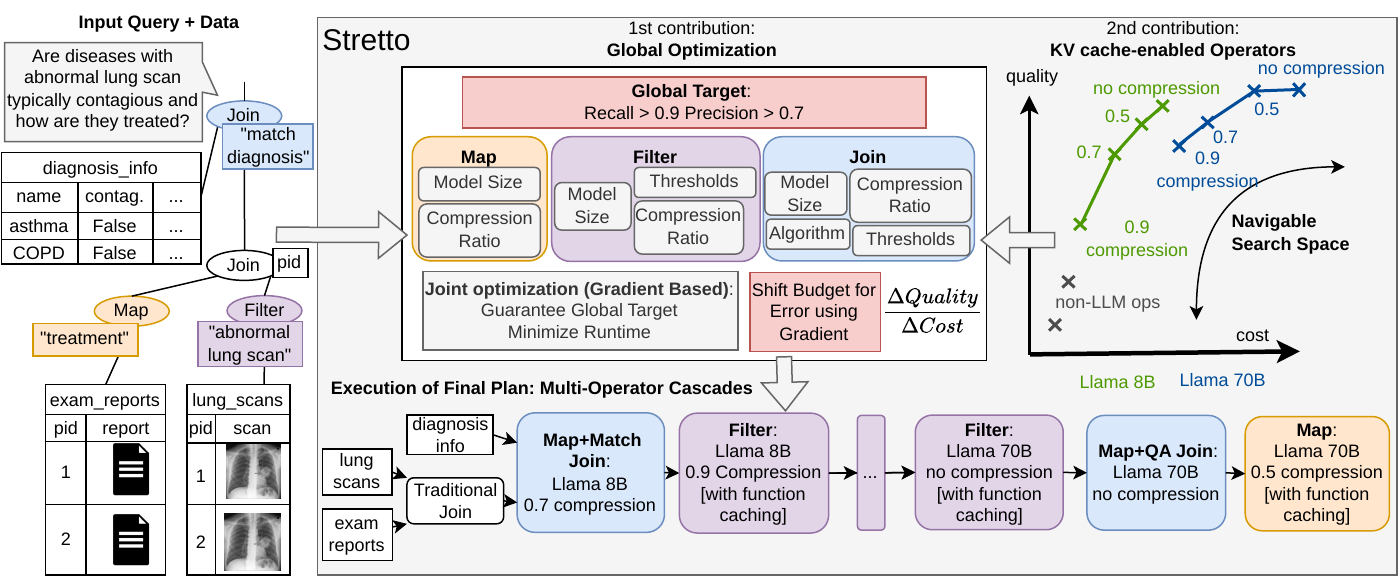}
  \vspace{-2ex}
  \caption{Overview of \ours{}. The input (left part) is a user query represented as a logical plan with semantic operators. The plan is globally optimized by selecting model sizes, KV caches, 
  and thresholds for each semantic operator. The optimizer (middle, upper part) allocates error budgets across operators to minimize runtime while meeting the global constraints regarding recall and precision. Our KV cache--enabled semantic operators, together with non-LLM-based options, span a search space of physical operators (right part) with clear cost--quality trade-offs. %
  The final physical plan (middle, lower part) is then executed. Traditional relational operations are handled by a classical engine (DuckDB in \ours{}) for optimization and execution.  %
  }
  \label{fig:overview}
  \vspace{-4ex}
\end{figure*}

Figure~\ref{fig:overview} summarizes \ours{}: given a query with classical and semantic operators, it globally optimizes the plan under quality constraints and executes the resulting physical plan.

\subsection{Problem Formulation}
\ours{} adopts a relational data model extended with multimodal data types~\cite{thalamusdb, lotusVLDB, UrbanB24}. 
As shown on the left of Figure~\ref{fig:overview}, we represent input queries as DAGs of relational operators (white) and semantic operators (color).
Each semantic operator has alternative physical implementations with different tunable parameters.
A semantic filter, for example, can be implemented either as an LLM-based operator or as a lightweight embedding-based one that thresholds the similarities between data and query embeddings.
Given a logical plan and a user specification of end-to-end quality constraints on the query output (e.g., precision and recall targets), the goal of \ours{} is to jointly select
(a) physical implementations for semantic operators, including different KV cache variants, and
(b) threshold parameters, such that the resulting execution plan satisfies the global quality constraint and minimizes query runtime.

\subsection{Optimization with Global Guarantees}

A first novel component of \ours{} is its query optimizer, illustrated in the middle part (top) of Figure~\ref{fig:overview} (\emph{Global Optimization}).
Rather than decomposing a global quality target into independent per-operator quality targets (e.g., as done by Lotus), \ours{} jointly optimizes all semantic operators in the plan.
Because operators exhibit heterogeneous cost--quality trade-offs that also depend on data and query, \ours{} employs a gradient-based optimization approach that dynamically allocates the allowable error budget across operators to minimize overall execution cost.
To enable this, we introduce a continuous relaxation of the search space that supports simultaneous optimization over categorical choices (e.g., model size selection) and continuous parameters (e.g., similarity or confidence thresholds).
This joint optimization is crucial: operator parameters, such as thresholds, directly influence the amount of error downstream operators must compensate for, thereby affecting which models are required to satisfy the global quality constraint.
The optimizer's output is an execution plan where, for each semantic operator, as shown in Figure~\ref{fig:overview} (bottom), a physical execution strategy (mode, thresholds, etc.) was chosen. 
\ours{} leaves the optimization of the classical relational part of the plan, such as picking physical operators for (non-semantic) joins, to a traditional engine. %
More details about the optimization %
are in Section \ref{sec:optimizer}.

\subsection{A New Physical Operator Layer}\label{sec:new-physical-layer}

Beyond global optimization, \ours{} introduces new physical execution strategies for semantic operators that exploit the idea of \emph{KV cache} at varying compression ratios (low to high) to implement different variants of semantic operators. %
These variants serve as the physical execution strategies our optimizer selects from, exposing runtime--accuracy trade-offs that allow the system to choose faster, highly-compressed operators when lower accuracy is requested, as illustrated on the right of Figure~\ref{fig:overview} (\emph{KV cache--enabled Operators}).
For KV cache compression, \ours{} uses a \emph{query-agnostic} compression technique based on \emph{Expected Attention}~\cite{expectedattention}.
Our technique identifies important tokens that are useful independently of a specific query, making the resulting compressed caches reusable across semantic operators.
Importantly, higher compression ratios also reduce memory consumption and enable larger batch sizes, yielding faster execution at the cost of reduced accuracy.
KV caches for semantic operators are constructed once per dataset (like materialized views) in an offline phase and can then be reused across all queries.
At query time, operators reuse these caches and skip prefill, leading to substantial speed-ups.
Precomputing KV caches comes with write amplification (i.e., additional storage needed when using KV cache operators vs. not using KV caches, which includes data and the models) of $3\times$ in our experimental setup,
but enables speedups of up to $22\times$ as shown in Section \ref{exp:ablation}.
We can also reduce storage by using an index-based scheme for KV cache storage, which can reduce storage overheads up to 44\% as discussed in Section
\ref{sec:constructing-caches}.
Thus, precomputed KV caches are best understood as a \emph{budgeted physical design choice}, analogous to deciding which indexes or materialized views to build under a storage budget: a deployment needs to only materialize the profiles that fit its budget and are useful for its workload.
In particular, uncompressed caches (i.e., Llama-3.1 70B and Llava-Next 72B without compression), which dominate storage, need not be materialized and can instead use standard on-the-fly inference.
As we report in our evaluation, we find that a small number of highly compressed KV caches already enable the bulk of possible speedups.

\subsection{A Navigable Search Space}
\label{sec:search-space}

A major difference to previous optimizers such as Abacus and Lotus is that \ours{} makes use of a %
navigable search space for semantic operators, allowing our optimizer to %
trade off query execution cost and result quality. %
Its search space spans four dimensions. %

\emph{(1) KV cache compression ratios.} For each model, \ours{} exposes multiple KV cache compression ratios, forming a ladder of physical operator variants with progressively lower runtime and accuracy. Higher compression reduces storage footprint and enables larger batch sizes, yielding faster execution.
\emph{(2) Model sizes.} \ours{} considers models of different sizes, providing coarse-grained cost--quality trade-offs: smaller models offer lower latency, while larger models achieve higher accuracy at increased computational cost.
\emph{(3) Non-LLM-based operators.} %
The search space includes inexpensive non-LLM operators, such as embedding-based filters, which can prune fractions of the data before invoking expensive models.
\emph{(4) Multi-stage operator cascades.} %
A single logical operator (e.g., a semantic filter) may be implemented as a cascade of physical operators.
With such cascades, cheaper semantic operators handle easy cases and filter most tuples, reducing the load on high-cost models.
Unlike Lotus~\cite{lotusVLDB}, however, \ours{} supports cascades with an arbitrary number of operators.

Unlike prior spaces based mainly on model and prompting variants, these dimensions expose comparatively structured cost–quality trade-offs that the optimizer can navigate.

%% file: sections/optimization.tex
\section{Optimization with Global Guarantees}
\label{sec:optimizer}

The aim of the optimizer of \ours{} is to select physical execution strategies from our search space for all semantic operators in the given logical plan that guarantee global quality constraints and minimize cost.
In this section, we first describe the guarantees formally provided by \ours{}, before we explain the main component that enables global optimization: the gradient-based optimization of semantic operators.
Afterwards, we show how we integrate this technique into our full optimization procedure, handling plans with classical and semantic operators.

\subsection{Formal Definition of Guarantees} \label{sec:guarantees}

\ours{} guarantees that it selects a physical plan $P_{\text{o}}$ that has at least a user-defined precision and recall when compared to a golden plan $P_{\text{g}}$ (or human labels, if available).
 $P_{\text{g}}$ implements all logical operators with their highest-quality implementation, and leads to the best overall quality.
\ours{}'s guarantees are global, meaning they apply to the entire plan, not individual operators.
They are statistical and hold when the dataset $D$ is large enough, with a user-defined high probability (e.g., 0.95).
To pick $P_{\text{o}}$, \ours{} samples a small subset $S \subset D$ i.i.d. uniformly at random, and tunes the parameters and model choices of $P_o$ until it can guarantee the targets, while minimizing cost.
More formally, let $\mathcal{D}$ be the underlying input data distribution, such that $S = \{x_i\}_{i=1}^N$ is drawn i.i.d.\ from $\mathcal{D}$, $x_i \sim \mathcal{D}$.

We denote the result set of executing a plan $P$ on a (large) dataset $D$ (whose records are also drawn from $\mathcal{D}$) as $P(D)$. %
Then we can define distribution-level %
$\mathrm{Precision}_{\mathcal{D}} = \lim_{|D| \to \infty} \frac{|P_{\text{o}}(D) \cap P_{\text{g}}(D)|}{|P_{\text{o}}(D)|}$ and
$\mathrm{Recall}_{\mathcal{D}} = \lim_{|D| \to \infty}
 \frac{|P_{\text{o}}(D) \cap P_{\text{g}}(D)|}{|P_{\text{g}}(D)|}$.
Since $P_{\text{o}}$ is selected based on a small sample $S$, these distribution-level metrics are uncertain at optimization time.
\ours{} provides \emph{guarantees} over this uncertainty with \textit{Bayesian} \textit{credible intervals}.
Given user-specified recall and precision targets $\mathcal{T}_{\text{R}}$ and $\mathcal{T}_{\text{P}}$, and credible levels $\alpha_\mathrm{Recall}$ and $\alpha_\mathrm{Precision}$, %
\ours{} ensures, under the statistical assumptions above, that it picks a plan s.t. %
$\Pr\!\left( \mathrm{Recall}_{\mathcal{D}} \ge \mathcal{T}_{\text{R}} \right) \ge \alpha_\mathrm{Recall}$ and
$\Pr\!\left( \mathrm{Precision}_{\mathcal{D}} \ge \mathcal{T}_{\text{P}} \right) \ge \alpha_\mathrm{Precision}$.
In other words, the targets are met with probability $\alpha_\mathrm{Recall}$ and $\alpha_\mathrm{Precision}$ (in our experiments, we set $\alpha_\mathrm{Recall} = \alpha_\mathrm{Precision} = 0.95$).
To do so, \ours{} tunes the plan's parameters and model choices s.t.\ the posterior lower bounds on precision %
and recall %
given the sample $S$ are larger than the targets: 
$\ell_{\alpha_\mathrm{Recall}} \ge \mathcal{T}_{\text{R}}$ and $\ell_{\alpha_\mathrm{Precision}} \ge \mathcal{T}_{\text{P}}$.
At execution time, we measure precision and recall on a finite dataset $D$.
The empirical precision and recall on $D$ will be very close to the true distribution level precision and recall if $D$ is large enough.
Importantly, there is no assumption on the size of $S$, but a larger sample size allows \ours{} to be more confident in using cheaper models.
This is a statistical guarantee at the distribution level relative to the golden plan $P_{\text{g}}$, not an adversarial worst-case guarantee on every finite dataset. To summarize:

\begin{proposition}[Posterior Guarantee on Plan Quality] \label{prop:system_guarantee}
Given a profiling sample $S$ drawn i.i.d.\ from $\mathcal{D}$, let $P_{\text{o}}$ be the plan selected by \ours{} such that its posterior lower bounds satisfy $\ell_{\alpha_\mathrm{Precision}} \ge \mathcal{T}_{\text{P}}$ and $\ell_{\alpha_\mathrm{Recall}} \ge \mathcal{T}_{\text{R}}$. 
Under the posterior induced by $S$,
\begin{align*}
    \Pr\!\left( \mathrm{Precision}_{\mathcal{D}} \ge \mathcal{T}_{\text{P}} \mid S \right) &\ge \alpha_\mathrm{Precision} \quad \text{and} \\ 
    \Pr\!\left( \mathrm{Recall}_{\mathcal{D}} \ge \mathcal{T}_{\text{R}} \mid S \right) &\ge \alpha_\mathrm{Recall}.
\end{align*}
That is, $P_{\text{o}}$ satisfies the user targets with the stated posterior probability at the distribution level, relative to the golden plan $P_{\text{g}}$.
\end{proposition}
Note that there is no guarantee on cost (i.e., runtime) in \ours{}.
Instead, the goal is to find the cheapest plan possible while simultaneously guaranteeing the precision and recall targets.
Thus, when targets are too strict, \ours{} naturally falls back to $P_{g}$.

\noindent\textbf{Discussion.}
To test if the golden plan $P_\text{g}$ is a reliable proxy, we evaluated 15 manually labeled queries. While $P_\text{g}$ tracks human labels closely, it occasionally incurs minor target violations when evaluated against ground truth. When \textsc{Stretto} uses human labels directly and finds a feasible plan, strict ground-truth guarantees hold. Thus, for high-stakes domains, we recommend optimizing against human labels. For other domains, $P_\text{g}$ is the better option to avoid manual profiling overhead. Full results are in Appendix~\ref{appendix:human-labels}.

\subsection{Gradient-Based Query Optimization}

The goal of \ours{}'s optimizer is to pick and parameterize the most cost-effective models for all semantic operators of a plan %
while guaranteeing user-defined targets on global (i.e.,
query-level) precision and recall.
This is a constrained optimization problem in which the goal is to minimize cost, subject to the quality targets.

\noindent\textbf{Optimization Challenges.}
While the search space introduced in Section \ref{sec:overview} allows the optimizer to adjust the trade-off between quality and runtime, it is not easy to search and find an optimal solution.
It combines continuous parameters, such as similarity thresholds for embedding-based semantic filter operators, with discrete choices, such as which model size to pick.
Therefore, purely combinatorial search procedures, such as Pareto-Cascades as used in Abacus \cite{abacusArxiv}, would require discretizing the continuous parameters, leading to either a large search space or limited granularity.
Moreover, as discussed before, the choice of a threshold for one physical implementation of an operator can directly affect the model choice for another logical operator, given the global nature of targets.
Therefore, optimizing the parameters locally per logical operator like Lotus~\cite{lotusVLDB} is also suboptimal.
Finally, assuming independence between semantic operators, as done by Abacus to avoid the exponential search space explosion, is also problematic, as we will show, since violations can lead to missing the quality targets.

\noindent\textbf{Intuition.}
When constrained by global targets on quality, we need to allocate the budget for query processing errors to individual operators in a way that allows us to select the most cost-effective implementations possible overall, rather than splitting the quality targets evenly.
If, for each logical operator, we can estimate how much additional cost \((\Delta \mathrm{Cost})\) is required to guarantee a small increase in quality \((\Delta \mathrm{Quality})\), and how much cost can be reduced by relaxing quality, then we can navigate the search space using the marginal trade-off \(\frac{\Delta \mathrm{Quality}}{\Delta \mathrm{Cost}}\), assigning stricter targets to operators for which improvements are cheapest.
Based on this intuition, we optimize query plans with semantic operators using \textit{gradient descent}.
Instead of enumerating different model combinations, we compute a gradient that indicates which models to replace to achieve the target and minimize cost.
However, gradient descent over a mixed discrete--continuous search space is not straightforward: discrete choices, such as which model to pick, must first be relaxed into continuous, differentiable ones.
Next, we discuss the overall optimization procedure and how gradient-based optimization is embedded in it; details of the gradient-based solution are then discussed in Section~\ref{sec:details}.

\subsection{Overview of Optimization Procedure}\label{sec:overview-optim}

\begin{figure}
  \centering
  \includegraphics[width=0.95\linewidth]{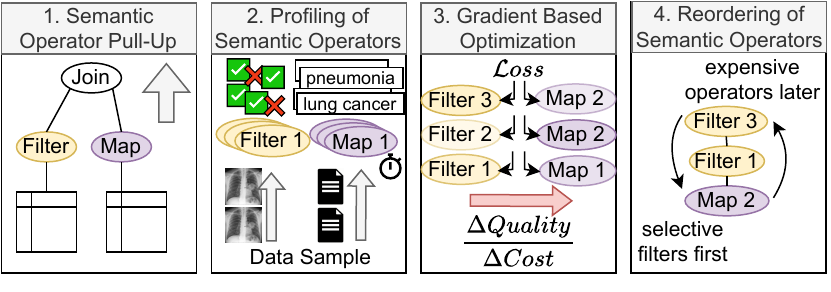}
  \vspace*{-2ex}
  \caption{The four steps of optimization in \ours{}. The optimizer (1) pulls up semantic operators (yellow and purple) above relational ones (white) to reduce expensive LLM calls, (2) profiles semantic operators on samples to estimate cost--quality tradeoffs, (3) applies gradient-based optimization over a continuous relaxation of operator and parameter search space with Bayesian precision/recall guarantees, and (4) reorders selected physical operators to minimize runtime.}
  \label{fig:optimization}
  \vspace*{-4ex}
\end{figure}

The gradient-based query optimization is embedded into a procedure that takes a logical plan consisting of relational and semantic operators and creates a physical execution plan, as shown in Figure \ref{fig:optimization} and explained in the following:

\noindent\textbf{Step 1. Semantic Operator Pull-Up.}
As a first step, to enable gradient-based optimization,
we pull up as many semantic operators as possible in the logical plan. %
The rationale is that semantic operators that call LLMs are orders of magnitude more expensive than relational ones.
By applying relational filters and joins first, %
traditional operators minimize the number of multi-modal items in the input of semantic operators
~\cite{horrila}.
This step groups semantic operators into what we call a pipeline: a sequence of semantic operators, on top of relational operators, that is then optimized as described next.
Semantic joins are executed by splitting them into a relational Cartesian operator and a semantic (but blocking) filter, which takes the join predicate. 
This strategy still allows the implementation of different techniques as part of the blocking semantic filter to reduce the join costs, as we discuss in the next section.
Note that we apply pull-up at the logical plan level, implicitly pulling up also cheaper implementations like embedding-based filters.
We do this to enable global optimization, which leads to large speedups as we show in the evaluation.
However, further optimizations, like pushing down such cheaper filter implementations again to reduce, e.g., the input cardinalities for traditional joins, are in principle possible after global optimization has finished.

\noindent\textbf{Step 2. Profiling Semantic Operators.}
As a second step, we profile the runtime/quality tradeoff of physical execution strategies for all semantic operators in the query plan, which is required as input for the optimization.
For this, we run all physical operators on the same small subset of data.
We sample tuples i.i.d.\ (100 in our experiments) from the pipeline input (i.e., the output of the traditional part) uniformly at random.
However, as a major difference to existing approaches, since we ultimately care about plan quality at the pipeline level, not per operator, it is not sufficient to record per-operator statistics during profiling.
Instead, we record the \emph{output} of each operator per tuple (i.e., the extracted values, computed embedding similarities, log-odds, depending on the operator).
That allows us to simulate many different pipeline configurations cheaply, i.e., different combinations of execution strategies for the semantic operators in a given query plan.
For instance, when there is a map and a filter operator in the pipeline, %
each combination of physical implementations might yield a different output and %
accuracy.
The gradient-based optimization %
cheaply simulates %
different combinations of physical implementations and cascades, whose pipeline-level outputs define the loss in Section \ref{sec:details} (see Equation \ref{eqn:soft_decision}).

\noindent\textbf{Step 3. Gradient-based Optimization.}
After profiling, we start gradient-based optimization. %
To make \ours{}'s search space differentiable, we must replace discrete parameters in the search space, such as discrete compression ratios, with continuous soft picks \cite{darts, diffml}.
That means the optimizer can temporarily select operators (e.g., one with a specific compression ratio) only partially, for instance 80\%.
This makes the search space smooth and differentiable, allowing the optimizer to take small steps along the gradient towards minimizing cost and guaranteeing the targets.
However, it is important that after optimization, we have a decision for each operator.
We therefore reduce smoothness during optimization by gradually lowering a temperature parameter, leading to a discrete decision on which operators to pick at the end.
To ensure that the user-specified precision and recall targets are met with high probability, we compute bounds on recall and precision (based on the profiling results) and optimize the model and parameter choices so that these lower bounds exceed the %
targets.
Since it would be invalid to optimize directly against frequentist confidence bounds (which could lead to p-hacking), we rely on Bayesian credible intervals.

\paragraph{Step 4. Reordering of Semantic Operators} 
Finally, we also optimize the structure of the query plans, because the order of operators is crucial: for instance, placing a cheap, selective filter at the beginning of the pipeline can massively reduce overall execution cost compared to placing it at the end.
However, optimizing the operator ordering with gradient descent is hard, for instance, because all operator positions are entangled, meaning moving one operator to the front must always result in moving other operators to the back, among other difficulties.
Therefore, we reorder the sequence of physical operator choices obtained from gradient optimization in a separate subsequent step.
We use dynamic programming, leveraging profiling results such as per-tuple processing time and selectivity, to determine the optimal order.
The details of operator reordering are explained in the Appendix.

%% file: sections/operators.tex
\begin{figure}
  \centering
  \includegraphics[width=\linewidth]{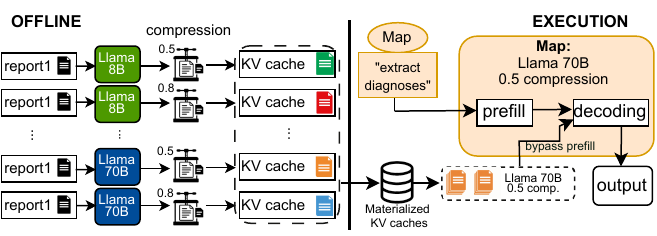}
  \vspace{-5ex}
  \caption{Offline, the KV cache for each item is precomputed and stored under profiles for different models and compression ratios. %
  At inference time, for a query (e.g., request to extract diagnoses), a profile is selected, the corresponding KV cache retrieved and its items batched %
  to the LLM for execution. As the KV caches have been computed, the prefill phase is bypassed, %
  significantly reducing latency and compute cost.}
  \label{fig:kvcache-overview}
  \vspace{-4ex}
\end{figure}

\vspace{2ex}
\section{KV cache--enabled Semantic Operators}
\label{sec:operators}
A second key building block of \ours{} is the KV-based physical execution strategies for semantic operators. \ours{} supports maps, filters, and joins using KV cache--enabled operators for two modalities: text documents and images\footnote{In principle, this approach can be used for any modality, including audio and others.}. 

\subsection{Constructing and Using Caches}
\label{sec:constructing-caches}

Executing a semantic operator requires encoding its text or image input, producing the model's KV cache. This \emph{prefill} step is computationally expensive and can dominate query latency. Since semantic data systems often know the underlying data at ingestion time, \ours{} precomputes KV caches offline and reuses them at query time, bypassing prefill, as illustrated in Figure~\ref{fig:kvcache-overview}.
For each dataset, \ours{} materializes compressed KV caches at different compression ratios, yielding reusable physical \emph{profiles} from which the optimizer selects at runtime. Crucially, cache construction is \emph{operator- and query-agnostic}, %
so the same profiles can be reused across semantic operators and future queries.

\paragraph{Offline: KV Cache Creation}
To compute the KV caches, we process all database images and text through the LLM and store the extracted caches on disk. Because future queries (e.g., specific filter predicates or map properties) are unknown at this stage, token compression must be \emph{operator- and query-agnostic}.
We adopt \emph{Expected Attention}~\cite{expectedattention} as an off-the-shelf technique to retain tokens by expected global attention, without needing a known prompt~\cite{CoralloP24}. 

Importantly, \ours{}'s contribution is orthogonal to the underlying compression algorithm. Rather than proposing a novel compression method, \ours{} integrates existing techniques by exposing KV caches at multiple compression ratios as distinct, navigable physical operators within a global query optimizer. This modular design allows substituting any state-of-the-art query-agnostic compression technique while enabling the optimizer to seamlessly trade cache size and query runtime against downstream accuracy.

No single cache profile is optimal across all semantic operators and workloads: simple filters tolerate aggressive compression, whereas extraction may need less compressed caches. We therefore maintain multiple cache \textit{profiles}, computed offline under different \textit{compression ratios}. Each has its own cost and quality characteristics, so \ours{}'s optimizer treats them as interchangeable physical operators and selects the cheapest ones satisfying the guarantees.
Keeping several profiles per logical operator is also \ours{}'s safeguard against the failure modes of any individual compression ratio \cite{liu2026semantic}: if a highly compressed profile yields poor estimated accuracy during profiling, the optimizer picks a less-compressed one instead.

Multiple profiles also introduce a storage trade-off, as the repository must fit within a disk budget; the key design choice is which ratios to retain. We curate, per modality, a small candidate set covering the relevant cost--quality space: we (i) evaluate a grid of \textit{model-ratio} profiles and measure quality and cost over query predicates randomly sampled from the dataset's attributes using an LLM, and (ii) prune offline both dominated profiles and those exceeding the storage budget.
This yields a compact repository that spans the trade-offs the optimizer explores. In our evaluation, modality dominates ratio selection: image workloads tolerate more aggressive compression because visual tokens are more spatially redundant than text.
Only profile \emph{retention} is workload-aware; compression itself remains query-agnostic, using only base-data activations before any predicate is known. We found that retention choices always generalized to unseen predicates in our case.

\paragraph{Online: Query Execution} 
When an optimized physical query plan is executed, the cache-based operators are applied to batches of images or texts by invoking an LLM over a precomputed (compressed) KV cache, rather than over the raw inputs. For this, at runtime, the operator first loads a collection of precomputed caches, appends an operator-specific query string derived from the semantic operator argument in the query plan (e.g., the filter predicate), and executes a single forward pass to produce one output per item. To maximize GPU utilization, we execute operators in batches using padded KV caches, with more aggressive compression enabling larger batch sizes and lower overall latency.
Each operator is defined by a cache profile, parametrized by a model (with its supported modality) and compression ratio. 
Caching the same images or texts under multiple profiles (e.g., different compression ratios and/or models) leads to many physical implementations of the same logical operator %
with explicit cost--quality trade-offs. %

\paragraph{Discussion}
Given the substantial runtime cost of semantic data systems, pre-materializing and compressing KV caches to speed up inference is a direction that must be studied. While this introduces storage overhead, the footprint can be constrained to a fixed budget by limiting the number of KV cache profiles generated and stored. This enables %
a manageable storage overhead, while yielding speedups of up to $22\times$. As an alternative to the full materialization of all profiles, we investigated an index-based storage strategy. %
Since highly compressed profiles share their KV pairs with the less compressed ones, they can be stored as lightweight index structures, and the highly compressed KV caches are reconstructed on-the-fly from a single physical baseline cache per model. Evaluated across the full set of profiles used for Figure~\ref{fig:kvcache-plot}, this saves \textasciitilde 21\% of storage for visual datasets (e.g., 830 GB saved for Artwork) and \textasciitilde 44\% for textual datasets (e.g., 156 GB saved for Movie), but incurs a runtime penalty (e.g., Movie: \textasciitilde 160s with indices on average, \textasciitilde 118s with physical caches, +36\% ; Artwork: \textasciitilde 131s versus \textasciitilde 79\,s, +66\%), since the reconstruction process requires loading and processing a larger-than-necessary physical baseline cache, increasing I/O overhead.

\subsection{Realizing Semantic Operators}

\paragraph{Semantic Filter and Map} A filter evaluates a Boolean predicate per item. 
Concretely, the model produces the logits for the tokens '1' and '0', which we use to classify whether to \emph{accept} or \emph{reject} the input tuple.
We then compute the log-odds between these tokens, which the system uses to classify the item as accepted, rejected, or unsure based on optimized thresholds.
A semantic map produces an output value per input item by synthesizing the requested fields. Compared to a filter, a map typically decodes longer outputs and is thus more sensitive to the compression ratio. 

\paragraph{Semantic Joins} For semantic joins, the KV cache layer provides two alternative physical implementations of the matching stage
with different cost--quality trade-offs.
The \emph{Extract+Match Join} and the \emph{Extract+QA Join} share an initial phase where a join-predicate-specific value is extracted from the pre-computed KV caches.
E.g., when joining patient reports with disease descriptions, the first phase extracts the diagnosis and the disease name from the respective KV caches.
The two algorithms differ in the second phase: the former filters pairs based on embedding similarity between the previously extracted values (e.g., diagnoses) to determine which pairs match, whereas the latter evaluates each combination directly via an LLM call, leading to more accurate decisions but higher latency.
Both reuse precomputed caches to support blocking-style execution, reducing the need for expensive pairwise LLM comparisons.

%% file: sections/details.tex
\section{Details of Gradient-based Optimization}
\label{sec:details}
\noindent In the following, we explain the details of gradient-based query optimization for semantic operators.
We start with a query that has only a single filter, which allows us to explain all important details, and then generalize to queries with multiple semantic operators. 

\subsection{Queries with a Single Semantic Filter} \label{sec:differentiable}
A single semantic filter in \ours{} can be implemented by a cascade of physical operators.
As physical operators in the cascade are sorted by their cost, which we know from profiling,
during the enumeration of different cascades, we only select or deselect physical operators, keeping their order intact.
More precisely, since we use gradient-based optimization, it is necessary that the decision about which operator to pick be continuous rather than binary.
Therefore, during enumeration, we actually enumerate mixtures of cascades, as we explain later.
But first, we begin with some preliminaries.

\paragraph{Tuple routing} 
First, we explain how a tuple is processed by a cascade, in which some operators are selected, others deselected, and the operators are sorted from least to most expensive.
The set of all available operators consists of $o_1$ (least expensive), ..., $o_n$ (most expensive).
Given an input tuple (e.g., containing an image), it will first be processed by the first selected operator, say, $o_1$ (if it is selected).
This physical operator can now \emph{accept}, \emph{reject}, or mark the tuple as \emph{unsure}.
When $o_1$ accepts or rejects a tuple, it will not be passed to the subsequent, more expensive operators.
Only when it is marked as unsure is it passed to the next more expensive operator in the cascade.
Thus, a tuple $t$ is accepted after operator $i$ if it was already accepted by a previous operator or if (1) all previous selected operators mark it as unsure, (2) $o_i$ is selected, and (3) it accepts the tuple.
For optimization (i.e., for gradient computation later), we formalize this behavior.
We define $\mathrm{accept}_{t, i}$, $\mathrm{reject}_{t, i}$, $\mathrm{unsure}_{t, i} \in \{0, 1\}$ to be 1 if a tuple $t$ is accepted, rejected or unsure respectively after physical operator $o_i$.
Then,
\vspace*{-2ex}

\begin{align}
    \mathrm{accept}_{t, i} &= \mathrm{accept}_{t, i-1} + \mathrm{unsure}_{t, i-1} \cdot \mathbf{1}^{selected}_{o_i} \cdot \mathbf{1}^{accept}_{o_i, t} \label{eqn:accept-i} \\
    \mathrm{reject}_{t, i} &= \mathrm{reject}_{t, i-1} + \mathrm{unsure}_{t, i-1} \cdot \mathbf{1}^{selected}_{o_i} \cdot \mathbf{1}^{reject}_{o_i, t} \label{eqn:reject-i} \\
    \mathrm{unsure}_{t, i} &= 1 - \mathrm{accept}_{t, i} - \mathrm{reject}_{t, i} \label{eqn:unsure-i}
\end{align}
\vspace*{-2ex}

where $\mathbf{1}^{selected}_{o_i} \in \{0,1\}$ is 1 iff $o_i$ is selected, and $\mathbf{1}^{accept}_{o_i, t}$, $\mathbf{1}^{reject}_{o_i, t}$ are 1 iff the operator $o_i$ accepts or rejects the tuple $t$.
Initially, all tuples are unsure: $\mathrm{unsure}_{t,0} = 1$, $\mathrm{accept}_{t,0} = \mathrm{reject}_{t,0} = 0$.

\paragraph{Estimating Cost (Runtime)} 
Next, we formalize our cost model.
The cost of the cascade is given by the per-tuple $cost_{o_i}$ of the individual physical operators $o_i$ and the number of tuples they process.
For a single tuple $t$, the cost of the cascade is given by  
\vspace*{-1ex}
\begin{align}
cost(t)=\sum_{i=1}^n \mathrm{unsure}_{t, i-1} \cdot cost_{o_i} \cdot \mathbf{1}^{selected}_{o_i} \label{eqn:cost}
\vspace*{-2ex}
\end{align} 

as an operator only needs to process unsure tuples.
Profiling gives the per-item cost of each operator implementation, from which we estimate the cascade's runtime on the full dataset.

\paragraph{Estimating Precision \& Recall} In addition to runtime, we also estimate and optimize the accuracy of a query plan with semantic operators. 
Next, we describe how we do this for a cascade of physical operators that implement a semantic filter. However, this generalizes to arbitrary pipelines with multiple semantic operators.
The precision and recall of a cascade $P$ are given by comparing the outputs to the golden cascade $P_\text{g}$.
The golden cascade $P_\text{g}$ uses only the most expensive operator (i.e., the largest LLM with the highest expected accuracy) for all input items.
When simulating a cascade during optimization and comparing it to its golden counterpart by using the recorded operator outputs on a data sample $S$ during profiling, we observe the cascade outputs $P(S)$ and $P_\text{g}(S)$.
Then we compute the true positives, false positives, and false negatives to compute the precision and recall as follows:
\begin{align}
TP_S &= \sum_{t \in S} \mathrm{accept}_{t, n} \cdot \mathbf{1}_{t \in P_\text{g}(S)} \label{eqn:sample-tp} \\
FP_S &= \sum_{t \in S} \mathrm{accept}_{t, n} \cdot (1- \mathbf{1}_{t \in P_\text{g}(S)}) \label{eqn:sample-fp} \\
FN_S &= \sum_{t \in S} (1-\mathrm{accept}_{t, n}) \cdot \mathbf{1}_{t \in P_\text{g}(S)} \label{eqn:sample-fn} 
\end{align}
\vspace*{-1ex}

Note that if we have human labels available, we replace $P_\text{g}$ in Equations \ref{eqn:sample-tp} - \ref{eqn:sample-fn} with the human labels, to obtain true positives, false positives and false negatives with respect to those.

\paragraph{Statistical Guarantees}
At optimization time, we need to try out different cascades and estimate their precision and recall on the full dataset.
The true positives, false positives, and false negatives observed on the sample allow us to compute the sample precision and recall.
However, because the sample is small, these metrics can change significantly on the full dataset.
Previous work \cite{spare-llm, lotusVLDB} used binomial confidence intervals, such as the Clopper-Pearson exact method \cite{clopper-pearson} and those based on the normal approximation \cite{lotusVLDB, supg}.
Unfortunately, %
as we optimize the cascades using gradient descent and try out many different cascades %
at each iteration, it is not valid to use such confidence intervals from frequentist statistics.

This is because each application of the confidence interval is equivalent to performing a statistical hypothesis test, and repeating such tests without correction leads to optimism bias (i.e., p-hacking).
Moreover, applying a correction, such as the Bonferroni correction~\cite{bonferroni}, yields very conservative bounds due to the large number of pipeline configurations that arise in an iterative optimization procedure like gradient descent.
Therefore, we take a different route.
Unlike frequentist confidence intervals, we optimize against Bayesian \textit{credible intervals}, since their definition is not based on hypothesis tests.
Instead, a 95\% credible interval for recall %
simply means we are 95\% sure the true recall value lies within the interval.

For recall, consider a random tuple from the result set of the golden cascade $P_\text{g}$ applied to the full dataset, $t \in P_\text{g}(D)$.
Now, the random variable that $t$ is also in the result set of plan $P$ is Bernoulli distributed $t \in P(D) \mid t \in P_\text{g}(D) \sim Bernoulli(\mathrm{Recall}_\mathcal{D})$, where the probability of success is given by the recall.
In Bayesian statistics, the model parameter (here, $\mathrm{Recall}_\mathcal{D}$) is uncertain, and its uncertainty shrinks as we observe data.
Using an uninformative, conjugate $\mathrm{Beta}(1,1)$ prior, and the observed $TP_S$, $FP_S$ and $FN_S$, we compute the posterior lower bounds $\ell_{\alpha_\mathrm{Recall}}$ and $\ell_{\alpha_\mathrm{Precision}}$ using the regularized incomplete beta function (derivation in the Appendix).

\noindent\textbf{Continuous Relaxation.}
Now that we can estimate precision and recall during optimization, we discuss how to find an optimal cascade by gradient descent.
To do so, we make the search space differentiable by replacing discrete choices with continuous ones.
We first move from discrete indicator functions $\mathbf{1}^{selected}_{o_i}$ to continuous parameters $\Sigma = \{\sigma_{o_i}  \in [0, 1]\}$ called \textit{pick factors}, and replace them in Equations \ref{eqn:accept-i} - \ref{eqn:cost}.
To ensure that they are in the range $[0, 1]$, we use a sigmoid function $\sigma_{o_i} = \mathrm{sigmoid_\tau}(s_{o_i})$, where $s_{o_i}$ is unconstrained and $\tau$ is the temperature that controls the smoothness.

\begin{figure}
  \centering
  \includegraphics[width=\linewidth]{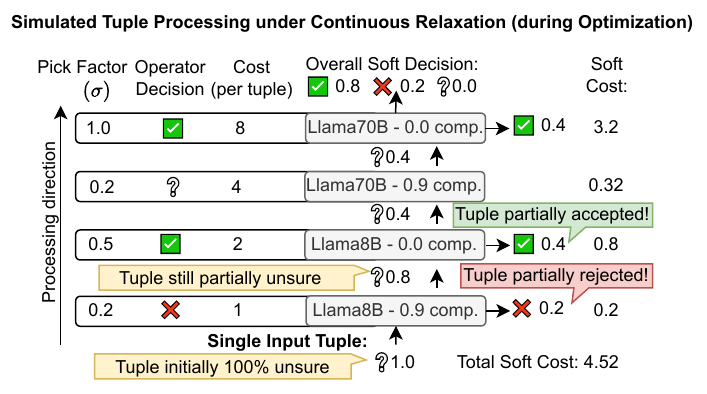}
  \vspace*{-4ex}
  \caption{
  The continuous relaxation allows the optimizer to temporarily explore mixtures of cascade configurations.
  In such a mixture of cascades, physical operators can be selected partially, see the pick factor $\sigma \in [0,1]$ on the left. Initially, the input tuple (at the bottom) is fully unsure.
  However, when it passes through the cascade, it is partially rejected, accepted, or marked as unsure, leading to a soft decision.}
  \label{fig:soft-pick}
  \vspace*{-2ex}
\end{figure}
If we pick $\tau=0$, we are again in the discrete, non-differentiable setting.
By replacing $\mathbf{1}^{selected}_{o_i}$ with $\sigma_{o_i}$, the values $\mathrm{accept}_{t, i}$, $\mathrm{reject}_{t, i}$, $\mathrm{unsure}_{t, i}$, as well as $TP_S, FP_S, FN_S$ become continuous, too.
Essentially, the optimizer can temporarily (i.e., during the optimization), explore mixtures of cascade configurations, see Figure \ref{fig:soft-pick}.
Initially, the input tuple (at the bottom) is fully unsure, but after passing it through the cascade, it becomes partially accepted/rejected, as each operator is partially selected.

For instance, assume the first operator in the cascade in Figure \ref{fig:soft-pick} (call it $o_1$), which uses Llama-3.1 8B at a compression ratio of 0.9, is 20\% selected ($\sigma_{o_1}=0.2$).
Since $o_1$ rejects the tuple, it becomes 20\% rejected, but stays 80\% unsure.
After passing it through the cascade, the tuple is accepted 80\% and rejected 20\% as later more expensive operators accepted the tuple.
Since $o_1$ was only partially selected, it contributed only partially to its per-tuple cost.
If the label accepts the tuple, it counts as $0.8~TP, 0.2~FN$, and $0.0~FP$, otherwise $0.0~TP,0.0~FN, 0.8~FP$.
Of course, we do not want operators to partially accept or reject a tuple during query execution.
Therefore, we gradually reduce the temperature $\tau$ over time using an exponential schedule and set it to $0$ after the optimization, when we extract the selected operators.

\paragraph{Optimizing Operator Selection}
Once the continuous search space is in place, we can use it to tune the pick scores.
For now, we assume that there is only a single logical filter in our query.
Thus, the precision and recall targets $\mathcal{T}_{\text{P}}$ and $\mathcal{T}_{\text{R}}$ directly apply to the filter.
Then it becomes a constrained optimization problem:
\begin{align}
   \Sigma_{optim} \in \argmin_{\Sigma} \sum_{t \in S} cost_{\Sigma}(t) \\
\text{s.t.} \;
\ell_{\alpha_\mathrm{Recall}} \geq \mathcal{T}_{\text{R}}, \quad \ell_{\alpha_\mathrm{Precision}} \geq \mathcal{T}_{\text{P}}
\end{align}

To optimize, we compute a loss $\mathcal{L}$ as follows:
\begin{align}
   \mathcal{L}_{cost} &= \frac{\sum_{t \in S} cost_{\Sigma}(t)}{|S| \cdot \sum_{i=1}^n cost_{o_i}} \label{eqn:cost-loss} \\
   \mathcal{L}_\mathrm{Recall} &=  ReLU(\mathcal{T}_{\text{R}} - \ell_{\alpha_\mathrm{Recall}}) \\
   \mathcal{L}_\mathrm{Precision} &=  ReLU(\mathcal{T}_{\text{P}} - \ell_{\alpha_\mathrm{Precision}}) \\
   \mathcal{L} &= \mathcal{L}_{cost} + \beta \mathcal{L}_\mathrm{Precision} + \beta\mathcal{L}_\mathrm{Recall} \label{eqn:loss}
\end{align}

Note that the loss on precision $\mathcal{L}_\mathrm{Precision}$ and recall $\mathcal{L}_\mathrm{Recall}$ are only active (i.e., gradient $\ne 0$) if the current cascade configuration violates the respective target (e.g., if $\mathcal{T}_{\text{R}} > \ell_{\alpha_\mathrm{Recall}}$).
Moreover, all loss components $\mathcal{L}_{cost}$, $\mathcal{L}_\mathrm{Precision}$, $\mathcal{L}_\mathrm{Recall}$ are in the range $[0,1]$.
We optimize the loss using the Adam optimizer \cite{adam}, while ensuring that the losses associated with the constraints are 0 after optimization to meet the overall quality criteria.
To do so, we run the same gradient descent optimization 256 times in parallel on the GPU, each with a different constraint penalty weight $\beta$ drawn from an exponentially spaced grid.
Then we pick the cheapest converged solution with zero constraint violations.
If we cannot find a solution with zero violations, we report it to the user and fall back to the golden plan.
In our experiments this happens when we optimize against human labels, where targets can be genuinely unreachable.

\paragraph{Optimizing Tunable Parameters}
Besides selecting operators for the cascade, \ours{} also tunes operator parameters, such as thresholds on embedding similarity or log-odds of output tokens.
To formalize that we define the set of tunable parameters $\Theta = \Theta_1 \cup \dots \cup \Theta_n$, where $\Theta_i$ are the tunable parameters of operator $o_i$.

We can integrate the optimization of such parameters into the above optimization problem by replacing $\mathbf{1}_{o_i, t}^{accept}$, $\mathbf{1}_{o_i,t}^{reject}$, $\mathbf{1}_{o_i,t}^{unsure}$ by soft operator decisions $\pi_{i, t}^{accept}(\Theta_i)$, $\pi_{i,t}^{reject}(\Theta_i)$ and $\pi_{i,t}^{unsure}(\Theta_i) \in [0, 1]$ that depend on the tunable parameters $\Theta$ and sum to 1: $\pi_{i,t}^{accept}(\Theta_i)$ +  $\pi_{i,t}^{reject}(\Theta_i) + \pi_{i,t}^{unsure}(\Theta_i) = 1$.
For instance, for the case of an embedding-based filter with two thresholds $\theta^+, \theta^-$, where we accept all tuples with $similarity_t > \theta^+$ and reject all tuples with $similarity_t < \theta^-$ and mark all tuples in between as unsure, we compute the soft operator decisions as:
\begin{align}
\begin{bmatrix}
\pi^{accept}_t \\
\pi^{reject}_t \\
\pi^{unsure}_t
\end{bmatrix} =
\mathrm{softmax}_{\tau}
\begin{bmatrix}
 \mathrm{similarity_t} - \theta^+\\
\theta^- - \mathrm{similarity_t} \\
0
\end{bmatrix}
\label{eqn:soft_decision}
\end{align}

Remember that we recorded the outputs (i.e., the similarities) of the embedding-based filter on the sample during profiling, enabling this computation.
Note that with this formulation, $\pi^{accept}$ is the largest when the similarity is larger than the upper threshold $\theta^+$, and $\pi^{reject}$ is the largest when the similarity is smaller than $\theta^-$. If it is in between, $\pi_{unsure}$ is the largest.
We use the same formula to learn thresholds on log-odds for our LLM-based filters.
$\tau$ is again the temperature that decreases during the course of optimization.

\subsection{Generalization of Optimization} \label{sec:multiple-filters}

\noindent So far, we optimized a single logical filter $O$, but a query may contain arbitrarily many semantic operators, which \ours{} pulls to the top to form a pipeline. We start with pipelines of 
semantic filters and then generalize.

\paragraph{Multiple Semantic Filters}
Recall the intuition that we do not want to split the precision and recall targets evenly across the logical operators.
Instead, we want to allow that some operators can make more mistakes while others make fewer.

One way to achieve this is to assume independent operators and optimize a lower bound for the entire pipeline quality by multiplying the individual operator lower bounds, e.g.,
$\ell_{\alpha_\mathrm{Recall}} = \ell_{1,\sqrt[m]{\alpha_\mathrm{Recall}}}\cdot \ell_{2,\sqrt[m]{\alpha_\mathrm{Recall}}}\cdots\ell_{m,\sqrt[m]{\alpha_\mathrm{Recall}}}$ where $m$ is the number of operators and  $\ell_{i,\sqrt[m]{\alpha_\mathrm{Recall}}}$  denotes the recall lower bound of the $i$-th logical filter with credible level $\sqrt[m]{\alpha}$.
While the independence assumption is often used in database optimization to simplify the optimization problem, it is problematic when reasoning about end-to-end guarantees.
Consider two consecutive filters applied to medical images: one selecting images with signs of pneumonia, and another selecting images with signs of lung cancer.
To obtain global performance guarantees, the pipeline must perform reliably on images that satisfy both conditions.
This property does not follow automatically from each filter performing well in isolation.
As \ours{} does not assume independence, we compute a global lower bound on recall and precision by comparing the output of the \emph{entire} plan %
with that of the golden plan %
on a sample. %
As in the single-filter case, we compute true positives $\mathrm{TP}_{global}$, false positives $\mathrm{FP}_{global}$, and false negatives $\mathrm{FN}_{global}$ of the full pipeline with multiple operators.
This allows us to compute and optimize a global lower bound for recall (and precision) using the regularized incomplete beta function as before. The formula is in the Appendix.
Crucially, this means \ours{} does not assume independent per-operator samples: during profiling, the \emph{same} sampled tuples are run through all candidate physical operators of all logical operators in the pipeline, and $\mathrm{TP}_{global}$, $\mathrm{FP}_{global}$, $\mathrm{FN}_{global}$ are computed from the joint pipeline output rather than aggregated from per-operator statistics. Equivalently, \ours{} treats the entire multi-operator pipeline as a single black-box Bernoulli event per sampled tuple (whether the pipeline's output for that tuple agrees with the golden pipeline), and the Beta posterior is applied at this pipeline level, not per operator. %
\ours{}'s guarantees are global rather than local: the optimizer adjusts per-operator dials to satisfy one end-to-end constraint.

\paragraph{Map Operators}
So far, we discussed filter operators.
We now turn to how \ours{} optimizes map operators in pipelines with other semantic operators.
A map operator transforms an input tuple by computing a new value and storing it in a new column.
For example, a map might extract a diagnosis from a patient's textual report and store it in a \emph{diagnosis} column.
For a given input tuple, different physical implementations of the map operator may produce different output values, e.g., different extracted diagnoses from the same report.
During profiling, we apply all candidate map operators to each tuple and collect all produced values.
For each value, this process yields a different potential output tuple for $t$.
From this perspective, choosing a particular map operator is equivalent to selecting a subset of these potential output tuples. %
Thus, map optimization can be treated analogously to filter optimization: instead of selecting which input tuples to keep, the map selects which output tuples to materialize. %
For instance, suppose there are two alternative physical operators, $o_1$ and $o_2$, that extract diagnoses from text: $o_1$ extracts ``fever'' (incorrect), while $o_2$ extracts ``coughing'' (correct). 
Choosing $o_2$ means accepting the tuple with coughing as the diagnosis and rejecting the one with fever using the gradient-based optimizer.

\paragraph{Join Operators}
Finally, from our optimizer's point of view, a semantic join is simply a (blocking) filter over the Cartesian product of input tuples, so that all previously introduced methods work seamlessly for the join as well.
Moreover, the choice of algorithm (Extract+QA and Extract+Match) is a separate discrete parameter of the join, optimized in the same way as other discrete parameters, such as the compression ratio.
The different algorithms resemble different physical operators that the optimizer can pick.

%% file: sections/experiments.tex
\begin{figure*}
  \centering
  \includegraphics[width=\linewidth]{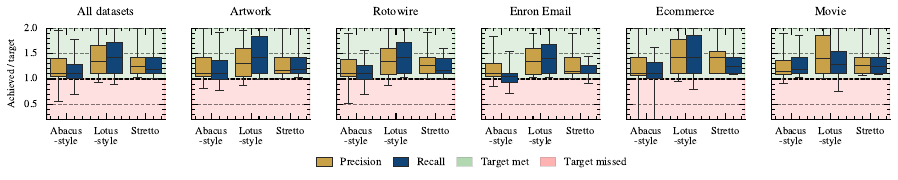}
  \includegraphics[width=0.9\linewidth]{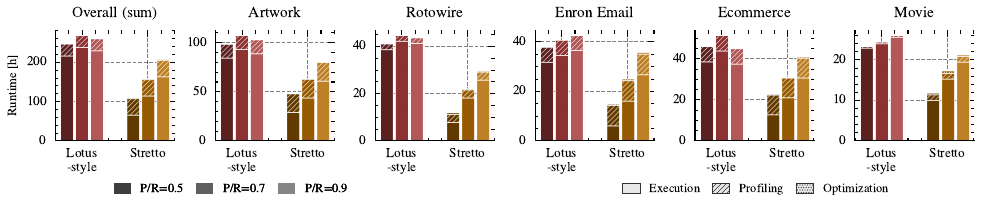}
  \vspace{-2ex}
  \caption{Runtime and Accuracy of \ours{}. Top (Accuracy): shows whether the global targets are met (\emph{Achieved / target} > 1). We show the distribution of all queries using a boxplot, with the whiskers at the 5th and 95th percentile, so that the lower whisker corresponds to the credible level (95\%). Thus, an approach meets its statistical guarantees when the entire boxplot is above the \emph{Achieved / target = 1.0} line.
  \ours{} is the most reliable in this regard, meeting the target overall.
  Bottom (Runtime): runtime comparison of the two optimizers that have statistical guarantees. Runtime includes profiling, optimization, and execution of all queries. \ours{} outperforms Lotus on all datasets.
  }
  \label{fig:guarantees-plot}
  \vspace{-2ex}
\end{figure*}

\vspace{2ex}
\section{Experimental Evaluation}
\label{sec:experiments}
In this section, we present the results of our experimental evaluation of \ours{}.
For our benchmark, we selected five datasets from previous work, along with a large number of queries. Beyond comparing \ours{} to prior systems (Exp 1), we isolate the two main sources of improvement, KV cache--enabled operators (Exp 2) and global optimization (Exp 3), and then study the storage, profiling, labelling, and ablation trade-offs of \ours{} (Exp 4--6).

\subsection{Experimental Setup}
\label{sec:experimental-setup}
\paragraph{Datasets} Our benchmark comprises five datasets.

\noindent\emph{(1) Artwork} \cite{UrbanB24} contains 1000 images depicting paintings from different historical periods, along with tabular information.

\noindent\emph{(2) Rotowire} \cite{rotowire,%
UrbanB24, eleet} contains 728 reports about basketball teams and players, along with natural-language match reports.

\noindent\emph{(3) Email} \cite{enron,%
palimpzestCIDR} consists of 1000 real-world corporate emails exchanged by employees of the Enron Corporation. It includes communications on 
topics such as business management decisions.

\noindent\emph{(4) Movie} \cite{sembench} contains text-only movie reviews, from which we selected 10,000 
to test how \ours{} works on larger dataset sizes.

\noindent\emph{(5) E-Commerce}
\cite{sembench} includes both images and text, enabling evaluation of queries that involve operators across modalities. It consists of images of items for sale, 
each accompanied by a textual description.
Again, 1000 images and their descriptions were selected randomly.

These datasets cover multiple modalities and a diverse range of domains to demonstrate \ours{}'s generality.

\paragraph{Queries}
The datasets contain too few queries to show that a system meets quality targets with high probability.
Hence, we generate 60 to 80 queries per dataset.
First, for each dataset, we manually create a natural language expression for 10 semantic filters and 10 semantic maps based on the original queries. For the Artwork and Movie datasets, which allow meaningful semantic joins, we additionally create 10 semantic join predicates.
Then, using 6 query templates per dataset, each containing 1-4 semantic operators, we generate queries by randomly inserting natural language expressions into the placeholders.
We randomly shuffle the semantic operators in the query and ensure that each resulting query is non-empty.

\paragraph{Operators}
\ours{} uses several physical operators across different KV cache profiles. As our smaller models, we use Llama-3.1 8B \cite{llama3} and Llava-Next 8B \cite{llava-next}, while for the larger models, we select the 70B/72B variants. 
If not mentioned otherwise in an experiment, we assume a strict storage budget.
Therefore, the optimizer can always choose from operators that use these models as-is, using standard LLM-inference.
In addition, the optimizer can also choose from two highly-compressed KV cache--enabled operators.
For the image modality, we add operators using Llava-Next 8B with a compression ratio of 0.9, and Llava-Next 72B with a compression ratio of 0.99.
In the text modality, we employ Llama-3.1 8B and Llama-3.1 70B, both with a compression ratio of 0.8.
The storage cost for these operators is 100 GB on the largest text dataset, Email (29 GB for the small model, 71 GB for the large one), and 88 GB on the largest image dataset, Artwork (30 GB for the small model, 58 GB for the large one). The numbers for the other datasets are in the Appendix.
As simple non-LLM operators, we use an image embedding-based filter based on Blip \cite{blip} and a simple Python operator that uses code to extract information from text.
The largest models without compression serve as golden models in our experiments.

\paragraph{Baselines}
We compare \ours{} against different baselines to show the impact of our main contributions in terms of both meeting user-specified quality targets and reducing execution runtime.
For our first experiment, we compare our optimizer against different baseline optimizers from the state-of-the-art: the Pareto-Cascades optimizer introduced by Abacus \cite{abacusArxiv} (which is the optimizer of Palimpzest) and the SupG \cite{supg} optimizer used by Lotus \cite{lotusVLDB}.
Note that Abacus lacks a mechanism to statistically guarantee quality targets and cannot tune continuous parameters, such as thresholds, so we fix these parameters at sensible defaults.
SupG (Lotus), on the other hand, can tune thresholds and guarantees per-operator targets.
Therefore, we split the global target evenly into per-operator targets for Lotus.
Moreover, Lotus supports cascades with only two operators: a smaller model first -- we pick the uncompressed 8B variants of Llama and Llava-Next -- followed by the golden models, which are closest to the original configuration of SupG.

For a fair comparison, we integrated both optimizers into our system, to isolate optimizer behavior and avoid confounding differences in storage, execution, or systems engineering. Moreover, other baselines, such as the original Palimpzest~\cite{palimpzestCIDR} system (without Abacus) and ThalamusDB~\cite{thalamusdb}, are not directly comparable as baselines because they do not expose an optimizer.

\paragraph{Metrics}
As a first metric, we use query runtime, which is dominated by the LLM calls of the semantic operators. 
We do not measure cost in terms of monetary expenditure, as all models used are open source and no LLM inference is performed via external API calls.
We evaluate \ours{} with open-weight local models rather than API-only models because our operator layer relies on direct access to model internals, in particular KV cache materialization, compression, and reuse, which are typically not exposed by closed API-based services (although \ours{} supports API-only models).

To measure quality, we compute the precision and recall of the query results after optimization %
against 
the results of $P_{\text{g}}$.
For each query, we compute an \emph{Achieved / target} metric for precision $\frac{\mathrm{Precision}}{\mathcal{T}_{\text{P}}}$ and recall $\frac{\mathrm{Recall}}{\mathcal{T}_{\text{R}}}$ where we divide the achieved metric by the target.
The metric is larger than one if the target is met.

\paragraph{Hardware}
We used NVIDIA A100 GPUs (80GB): 4 for single-modality datasets, 7 for multi-modal ones.

\subsection{Exp 1: Accuracy and Runtime}
\label{exp1}\label{exp:accuracy-runtime}
In this experiment, we show that \ours{} is the only state-of-the-art method that reliably provides global quality guarantees. 
Moreover, we show that by leveraging global optimization at the plan level, \ours{} finds faster plans than the baselines.
To do so, we execute all queries on the five datasets, three times with different precision and recall targets, ranging from relaxed to strict targets.
The most relaxed target is 50\% for both precision and recall, the medium target is 70\% precision/recall, and the strict target is 90\% precision/recall.
We set $\alpha_\mathrm{Precision} = \alpha_\mathrm{Recall} = 0.95$, i.e., we require each target to be met with 95\% posterior probability for both Lotus and \ours{}.
For profiling, we uniformly sample 100 tuples for all approaches to isolate optimization, corresponding to 1\% of the Movie reviews and 10\% of the other datasets.
Abacus lacks a mechanism to statistically guarantee quality targets.
Therefore, we pick the plan from the final Pareto frontier that satisfies the targets on the sample and minimizes cost.

\begin{figure}
  \centering
  \includegraphics[width=1\linewidth]{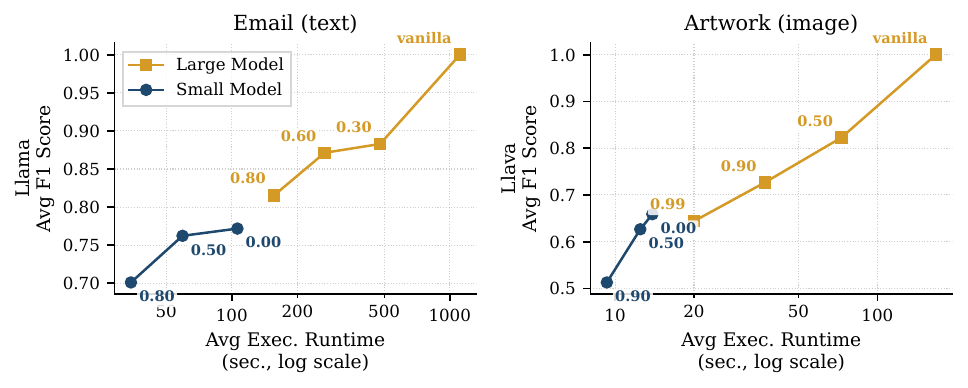}
    \vspace{-4ex}
 \caption{Runtime--quality trade-off for KV cache profiles, averaged over 20 single-operator queries on the Artwork and Email datasets. Points denote compression ratios: aggressive compression yields faster execution but lower quality. \textit{Vanilla} indicates standard large-model inference.}
 \label{fig:kvcache-plot}
  \vspace{-4ex}
\end{figure}

\paragraph{Accuracy} In Figure \ref{fig:guarantees-plot} (top), we show the distribution of the \emph{Achieved / target} metric across all queries using boxplots.
Importantly, we configured the lower and upper whisker ends to show the 5th and the 95th percentile.
Thus, a method is successful in achieving its statistical guarantees if the entire boxplot, including whiskers, is above one.
As shown by the left plot, where we plot all the queries over all datasets combined, \ours{} is the only method that can provide robust statistical guarantees. 
The ends of the lower whiskers are almost exactly on the \emph{Achieved / target} = 1.0 line, indicating that \ours{} can precisely tune the queries to meet the targets.
Note also that the interquartile range of the boxplot is very close to the \emph{Achieved / target} = 1.0 line, indicating that \ours{} is successfully trading off some quality for runtime while still meeting the targets.

Different from \ours{}, the baselines do not meet the targets.
Abacus' algorithm cannot statistically guarantee targets.
With a small sample, many of its queries miss them. %
Lotus' optimizer \cite{supg}, on the other hand, should meet the targets as we split global to local targets by assuming independence.
However, as we see, it does not consistently meet the recall and precision targets, as 7.6\% and 7.1\% of queries do not meet them.
This is due to the fact that we need to assume independence when splitting the target, which can result in target violations as discussed in Section \ref{sec:multiple-filters}.

 \noindent\textbf{Runtime.} We now compare runtime as shown in Figure~\ref{fig:guarantees-plot} (bottom), but leave out Abacus as it violates the targets significantly. \ours{} improves clearly over Lotus-style optimization, on every dataset and by $1.6\times$ on average, with the largest gain of $2.0\times$ on Rotowire. The reason is that Lotus only tunes the smaller model's threshold and cannot navigate our rich search space. It also tunes the operators only locally, requiring the global target to be split evenly into per-operator targets. \ours{}, in contrast, selects between different operators with different compression, can combine different operators, and uses cheap operators such as embedding-based filters or generated Python code. It also reorders and pulls apart the operator cascades for a single semantic operator, executing cheap pre-filters first and passing only those tuples deemed unsure to the more expensive operators.
 \ours{} has to profile a larger search space (including KV cache--based operators), so profiling takes 36\% longer than for Lotus. This cost is included in the runtime of Figure~\ref{fig:guarantees-plot} and is negligible compared to the execution time gains.
 Similarly, \ours{}'s optimization cost of about 30 seconds per query hardly matters in comparison to query execution gains.
 Regarding Lotus's local optimization, see also Experiment 3 for a controlled comparison of global versus local optimization.

\subsection{Exp 2: KV cache--enabled Operators}
\label{exp2}
One important tool for making our optimizer effective is the navigable search space provided by the KV cache compression strategy, which trades off quality and cost and affects the optimization.

\begin{figure}
  \centering
  \includegraphics[width=0.4\linewidth]{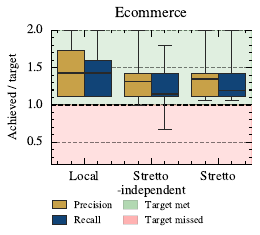}
  \includegraphics[width=0.4\linewidth]{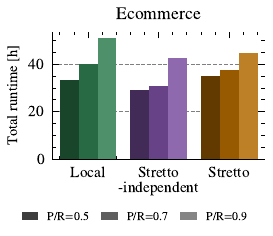}
  \vspace{-2ex}
 \caption{
 The effect of global vs. local optimization.
 Evenly splitting the global target into per-operator targets results in higher costs, assuming independence leads to faster plans, but no longer guarantees the quality targets.
 Global optimization meets the quality targets and finds fast plans.
 }
 \label{fig:meets-target-vs-local}
 \vspace{-2ex}
\end{figure}

\begin{figure*}
  \centering
  \includegraphics[width=\textwidth]{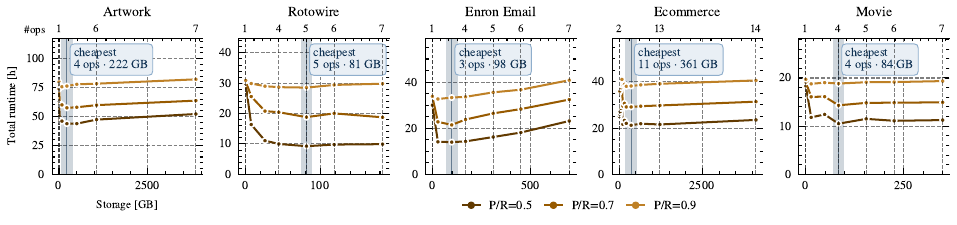}
  \vspace{-7ex}
 \caption{
 The storage/runtime trade-off. Adding KV cache--enabled operators increases optimizer flexibility but raises storage and profiling costs. Every plot starts left with \ours{} using no KV cache--enabled operators; i.e., only the biggest model standard inference, as well as cheap embedding- and Python-based operators. %
 Moving right, we add KV cache operators from those in Section \ref{exp2}, by lowest storage footprint. Adding operators leads to a sharp decline in total runtime, which is later countered by increased profiling cost. Usually, a few low-storage operators yield best speedup.
 E-Commerce requires 11 operators for its minimum, as the greedy enumeration strategy will first add many text operators before an image operator is added.
 }
 \label{fig:storage-tradeoff}
 \vspace{-2ex}
\end{figure*}

\begin{figure}
  \centering
  \includegraphics[width=0.95\linewidth]{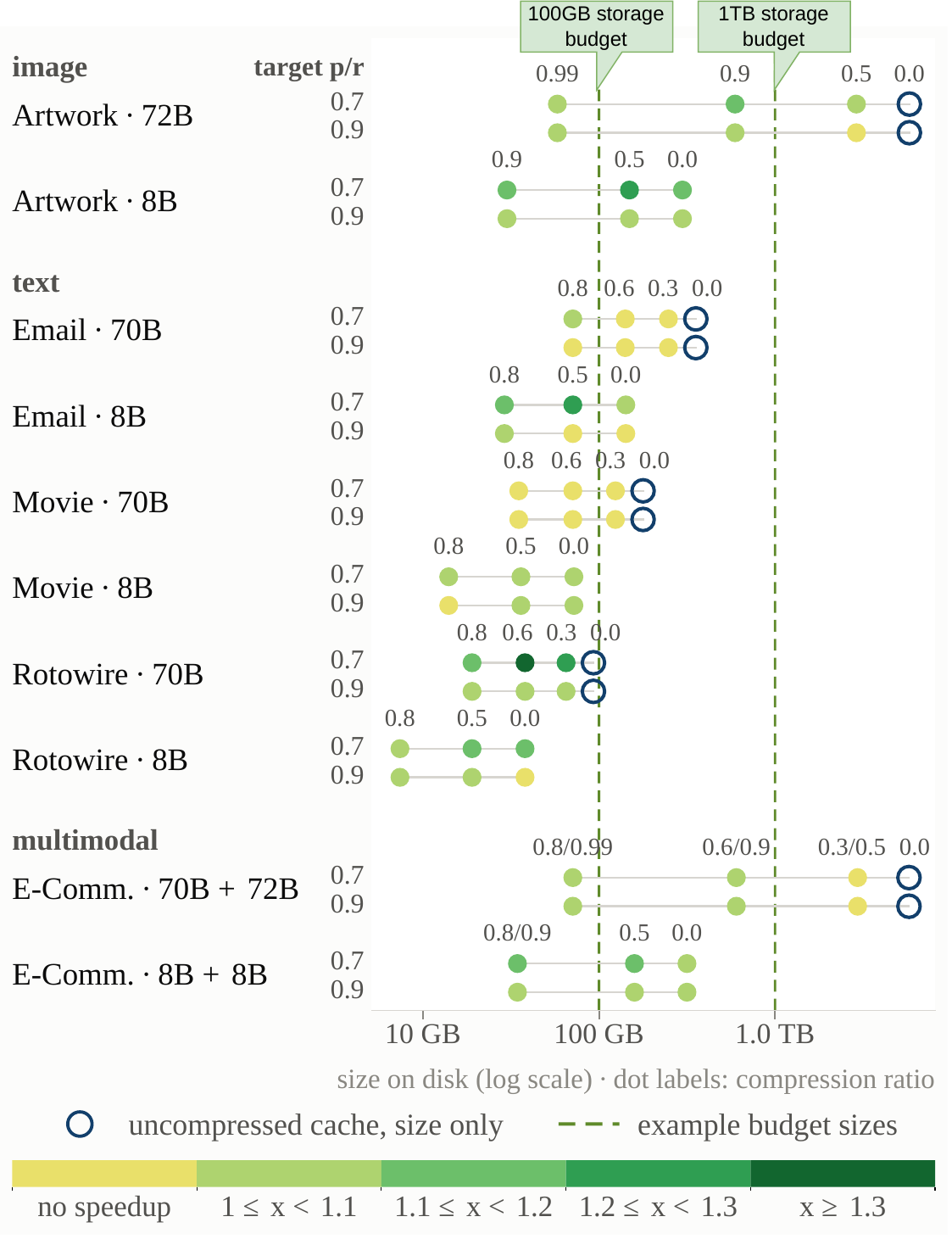}
  \vspace*{-2ex}
 \caption{KV cache footprint vs. end-to-end speedup across models and targets. Points represent isolated profiles available for the \ours{} optimizer. %
 }
 \label{fig:memory-vs-speedup-kvops}
 \vspace*{-4ex}
\end{figure}

\paragraph{Cost-Quality Trade-Off with KV-compression}
As shown in Figure~\ref{fig:kvcache-plot}, quality improves when increasing the model size and reducing the compression ratio, at the cost of higher execution time.
The compression ratio can be more aggressive in visual datasets, which involve images rather than text, due to the larger context and the higher proportion of background tokens that are not essential for completing the task.
Overall, these trends are consistent across tasks and modalities, allowing us to determine which configurations should be retained and which can be pruned during the pre-computation phase. We verified that these trends generalize across different model architectures as well, observing similar curves for the Mistral and Qwen families.
In Figure~\ref{fig:kvcache-plot},
we report one dataset from the visual modality (Artwork) and one from the text modality (Email) to illustrate that the expected trends are consistent across both modalities. In both cases, the reported F1-score and execution runtime are averaged over 20 single-operator query predicates sampled from the dataset's attributes, 10 with a semantic filter and 10 with a semantic map, drawn from the same pool used in the multi-operator queries of the experiment in Section~\ref{exp1}.

\subsection{Exp 3: Global vs. Local Optimization}
\label{exp3}
We now show the importance of global optimization by comparing \ours{} with two ablated variants that optimize locally per operator. Both use gradient-based optimization and all KV cache--enabled operators of Exp 2; only the loss computation differs.
\ours{}-Local splits the global target equally into per-operator targets (e.g., a recall target of 0.81 over two operators becomes 0.9 each).
\ours{}-Independent assumes operator independence and can therefore shift the error budget around: it treats each logical operator as a Bernoulli event and requires the product of the per-operator precision and recall lower bounds to exceed the target, with each bound being a Beta posterior quantile at credible level $\alpha^{1/k}$ for $k$ operators, so the product holds at level $\alpha$.
As expected, splitting targets naively is expensive: across all five datasets, it costs 19\% more execution time than \ours{} overall.
The effect on quality is best seen on E-Commerce (Figure~\ref{fig:meets-target-vs-local}).
Interestingly, assuming independence can find faster plans for some queries, e.g., $1.3\times$ faster in execution on E-Commerce, but at the price of more queries missing the targets.
Across all five datasets (not shown due to space constraints), the independence assumption misses at least one target on three of them. Global quality guarantees thus require global optimization.

\subsection{Exp 4: Storage/Runtime Trade-off}
\label{exp:storage-tradeoff}

\noindent\textbf{Manageable and  Beneficial Profiles.}
KV cache--enabled operators %
add storage cost.
To evaluate in which settings KV cache--enabled operators are manageable in terms of storage size and beneficial in terms of speedups, we plot them in Figure~\ref{fig:memory-vs-speedup-kvops}.
It maps the KV cache disk footprint (x-axis) against speedup (color-coded) for each dataset and model (y-axis).
Vertical dashed lines mark potential storage budgets.
To isolate individual gains, we augment a baseline search space (vanilla 8B and 70B models plus embedding and Python operators) with one KV cache profile at a time.
We calculate the end-to-end speedup per query (the ratio of baseline to augmented search space runtime) and report the geometric mean across all queries.
Precomputed KV caches benefit token-heavy datasets (e.g., Rotowire, Artwork) most.
Movie's very short texts yield limited gains, showing speedups only with 8B caches. %
Overall, 8B models and aggressive compression provide the best storage/performance trade-off.
Since we add one profile at a time, these gains are conservative relative to the full \textsc{Stretto} search space.

\noindent\textbf{Trade-off for Multiple Profiles}
Adding more operators increases both storage and profiling cost.
We start with \ours{} using only the largest models with standard inference plus embedding- and Python-based operators, and add KV cache--enabled operators from Section~\ref{exp2} in increasing storage-footprint order (Figure~\ref{fig:storage-tradeoff}).
Across datasets, configurations using 26 to 114\,GB already capture 80\% to 100\% of the obtainable end-to-end speedup, although the absolute minimum-runtime point may require more storage.
Further operators yield diminishing execution-time gains while profiling time and storage keep growing.
E-Commerce needs more operators because size-ordered enumeration adds several text profiles before the first highly compressed image profile.

\subsection{Exp 5: Effect of Sample Size}
\label{exp:profiling-tradeoff}
\begin{figure}
  \centering
  \includegraphics[width=0.56\columnwidth]{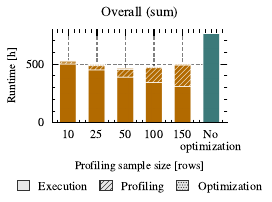}
  \includegraphics[width=0.43\columnwidth]{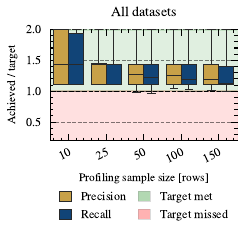}
    \vspace{-5ex}
 \caption{Profiling sample size effects. Left: larger samples increase profiling time linearly but decrease execution time, enabling the optimizer to confidently select cheaper operators. Right: \ours{} meets targets regardless of sample size.}
 \label{fig:sample-size}
  \vspace{-4ex}
\end{figure}

In this experiment, we vary how many tuples the optimizer profiles each operator on, from 10 to 150. We report the aggregated runtime and \emph{Achieved / target} distribution in Figure \ref{fig:sample-size} over all datasets and accuracy targets. %
In Figure \ref{fig:sample-size} (left), we see that a larger sample comes with more profiling costs.
At the same time, it tightens the confidence bounds the optimizer reasons about, so it can be more confident in using cheaper operators: execution time falls by about $1.6\times$ when increasing the sample size from 10 to 150.
Overall, we find that the best sample size is around 50 tuples, as it allows the optimizer to already pick some cheaper alternative implementations without spending 
too much time 
profiling.
For larger datasets, a larger sample size might be beneficial, as execution time increases and profiling time becomes less relevant.
Regarding the quality targets (Figure \ref{fig:sample-size}, right), \ours{} consistently meets them.

\subsection{Exp 6: Ablation Study}
\label{exp:ablation}
\ours{} has two main components: an optimizer that chooses, tunes and reorders the operators, and a search space containing KV-compressed operators.
To show the effectiveness of each, we start without either and add them one at a time to observe the execution speedups of each.
The results are shown in Table~\ref{tab:ablation}.
The starting point runs the best available operator on every tuple.
Adding the optimizer leads to large speedups across the board.
On average over all targets, we observe a speedup of more than $3\times$, rising to $5.3\times$ for the most relaxed quality target (and up to $32\times$ for individual queries).
Adding the compressed operators leads to additional speedups of up to $22\times$.
KV-operators shine when the targets leave room for them to be used, e.g., on the relaxed target we see more than $60\%$ of queries with improvement.
Note that the setting without KV-operators already uses cheap pre-filters like embedding-based filters and we still can further improve with better compression based filters.
On average, we see a large mean speedup of $1.48\times$.
With very strict targets, the opportunity for speedup decreases, however even for the strictest target we still observe speedups of up to almost $7\times$.
Regarding datasets, we see that speedups are more prominent with image datasets, as it is more expensive to compute the KV caches for those and skipping prefill therefore gives great benefits.

%% file: sections/related.tex
\section{Related Work}
\label{sec:related}
\paragraph{Guarantees for approximate query processing}
Providing statistical guarantees under sampling has a long history in approximate query processing~\cite{AQP,BlinkDB}. 
However, quality metrics in LLM-native pipelines (e.g., precision/recall) differ from numeric aggregation error and
our work targets \emph{task-level} quality metrics for pipelines to integrate guarantees into physical plan selection.

\paragraph{Cost--Quality Optimization for Semantic Operators}
\emph{Semantic operators} are the LLM-powered analogues of relational operators whose behavior is specified by natural-language parameters~\cite{palimpzestCIDR,UrbanB24}. 
Lotus %
develops optimization techniques (approximation cascades) %
to reduce cost while providing {statistical accuracy guarantees} at the granularity of individual operators \cite{lotusVLDB}.
Abacus's \emph{cost-based optimizer} for semantic operators searches a large implementation space using a small set of validation examples, enabling constrained optimization objectives over quality, cost, and latency \cite{abacusArxiv}, but it %
lacks statistical guarantees. %
Galois investigates %
logical/physical optimizations tailored to LLM retrieval calls, without semantic operators or statistical guarantees~\cite{galois-sigmod}.
MOAR introduces an agentic optimizer for semantic operator pipelines, focusing on logical/prompt-level pipeline rewrites for cost/accuracy optimization~\cite{wei2026multiobjective}. Semantic Histograms estimate visual-filter selectivity~\cite{urban2026selectivity}.
In contrast, we target \emph{end-to-end} %
guarantees on the \emph{pipeline output} and expand the physical design space by exposing KV cache-aware operators.

\begin{table}[t]
  \caption{Execution speedup by guarantee target for both ablation steps. Both steps yield large benefits, especially the optimizer. The KV-operators add further speedups of up to $22\times$ when the targets leave enough room for them to be used.} \label{tab:ablation}
  \vspace{-2ex}
  \centering
  \small
  \begin{tabular}{lrrrrrrr}
    \toprule
    Target & Mean & p90 & Max & >1.01$\times$ & >1.5$\times$ & >2$\times$ \\
    \midrule
    \multicolumn{8}{l}{\textit{Optimizer \textperiodcentered{} no optimization $\rightarrow$ \ours{}, wo KV cache--enabled ops}} \\
    P/R 0.5 & 5.34 & 11.10 & 32.09 & 92\% & 84\% & 78\% \\
    P/R 0.7 & 2.60 & 4.94 & 13.24 & 83\% & 66\% & 51\% \\
    P/R 0.9 & 1.71 & 2.76 & 9.41 & 72\% & 41\% & 25\% \\
    All targets & 3.21 & 6.56 & 32.09 & 83\% & 64\% & 51\% \\
    \multicolumn{8}{l}{\textit{Compression \textperiodcentered{} \ours{}, wo KV cache--enabled ops $\rightarrow$ full \ours{}}} \\
    P/R 0.5 & 1.91 & 3.32 & 22.85 & 62\% & 40\% & 26\% \\
    P/R 0.7 & 1.47 & 2.12 & 10.97 & 42\% & 21\% & 12\% \\
    P/R 0.9 & 1.07 & 1.11 & 6.89 & 14\% & 5\% & 1\% \\
    All targets & 1.48 & 2.31 & 22.85 & 39\% & 22\% & 13\% \\
    \bottomrule
  \end{tabular}
  \vspace{-4ex}
\end{table}

\paragraph{KV Cache Management and Compression}
KV caching is central to autoregressive decoding. Extensive work reduces its storage/time footprint via quantization, eviction, and token pruning while maintaining accuracy \cite{snapkv,CoralloP24,expectedattention,Corallo26}.
These works primarily target \emph{single-request} long-context inference efficiency, while we treat KV caches as \emph{materialized physical representations of base items} reusable across operators and queries; we incorporate (offline) compression ladders in the %
optimizer's search space as a cost--quality knob.

LLM serving systems like \textsc{vLLM} optimize transformer inference through batching, memory management, and KV cache reuse~\cite{vllm,liu2025optimizing}. 
These systems focus on single-query (or batched) inference and do not treat KV caches as a persistent physical representation. %

\paragraph{Model Cascades and Cost-Aware LLM Serving}
Finally, cascades that route ``easy'' inputs to cheap models and defer unsure inputs to more expensive models are widely used to reduce inference cost while retaining accuracy \cite{chen2024frugalgpt,routellm}.
Semantic operator systems also adopt cascades to trade cost for quality \cite{lotusVLDB}.
We adopt and generalize this idea by allowing multi-stage combinations of physical
implementations (including different cache variants) and by selecting thresholds
and routing policies with plan-level decisions.

%% file: sections/conclusion.tex
\section{Conclusion and Future Work}
\label{sec:concl}
We presented \ours{}, an execution engine for semantic operator pipelines under global quality constraints. Its gradient-based optimizer jointly selects physical operators and allocates error budgets to minimize cost while satisfying end-to-end quality targets. \ours{} exposes compressed KV caches as reusable physical representations, enabling fine-grained cost--quality trade-offs without repeated prefill. %
We show that \ours{} outperforms existing systems, and that exposing inference internals, such as the KV cache, to the query optimizer is essential for scaling LLM-native data processing.
Future work includes extending \ours{} to additional modalities and operators, such as semantic aggregation. We also plan to explore cluster-level KV caches to reduce storage growth, more cost-based semantic-operator placement (e.g., selectively pushing down cheap embedding filters), and adaptive profiling strategies that allocate additional samples when uncertainty remains high.

%% file: sections/appendix.tex
\newpage

\section{Appendix}

\subsection{Operator Reordering}
\label{app:reordering}

Algorithm~\ref{alg:dp-reordering} details the dynamic programming approach used by \ours{} to determine the optimal execution order of the selected physical operators. This procedure minimizes the total execution runtime by explicitly tracking the number of tuples remaining for each logical operator, applying the respective intra- and inter-operator selectivities at each state transition to accurately estimate costs.

After the optimizer selects a physical implementation for each operator, we additionally choose an execution order for the resulting physical operators to minimize runtime. 
We perform reordering after physical operator selection, since it is very hard to make the operator ordering differentiable and optimizable via gradient descent.
As a general principle, we prefer to execute operators that are both cheap and highly filtering early, so that fewer tuples reach later operators. Conversely, very expensive operators (e.g., large models without compression) should run late. In traditional query optimization, a common heuristic is to sort operators by $\frac{\mathrm{cost}(o)}{1-\mathrm{sel}(o)}$.
However, \ours{}'s execution model is not binary: a physical operator can \emph{reject}, \emph{accept}, or mark a tuple as \emph{unsure}.
Moreover, each logical operator can be implemented as a cascade of physical operators.
If a cheap stage accepts a tuple, the tuple can skip the remaining (more expensive) stages of the \emph{same} logical operator.
In contrast, tuples that are accepted (or unsure) must still be processed by physical operators belonging to \emph{other} logical operators in the plan. If a tuple is marked as unsure, it must also be processed by subsequent stages of the \emph{same} logical operator.

Therefore, each physical operator has two relevant selectivities.
We define the \emph{inter-operator} selectivity as the fraction of tuples not rejected by $o$: $sel^{inter}_o=\frac{\sum_{t\in S}\left(\mathbf{1}^{accept}_{o,t}+\mathbf{1}^{unsure}_{o,t}\right)}{|S|}$,
which corresponds to the tuples that subsequent \emph{different} logical operators still need to process.
We also define the \emph{intra-operator} selectivity as the fraction of tuples that remain unresolved within the cascade:
$sel^{intra}_o=\frac{\sum_{t\in S}\mathbf{1}^{unsure}_{o,t}}{|S|}$,
which corresponds to the tuples that subsequent physical operators for the \emph{same} logical operator must process.
Both selectivities can be estimated via simulation from the profiling results after the pipeline configuration is fixed.

To determine the optimal execution order, we use \textit{dynamic programming} (DP) and explicitly track the remaining tuples per logical operator.
Let $O = \{O_1, \dots, O_n\}$ denote the set of logical operators, and let $o = \{o_1, \dots, o_m\}$ denote the set of physical operators.
Each physical operator $o_i$ implements exactly one logical operator, denoted by $impl(o_i) \in O$.
For each subset of physical operators $\mathcal{S} \subseteq o$, we define the dynamic programming state
\[
DP[\mathcal{S}] = (C_\mathcal{S}, N_1^\mathcal{S}, \dots, N_n^\mathcal{S}),
\]
where $C_\mathcal{S}$ is the minimum total execution cost for executing exactly the operators in $\mathcal{S}$, and
$N_i^\mathcal{S}$ is the number of tuples remaining to be processed by logical operator $O_i$ after executing $\mathcal{S}$.

The initial state is defined as
\[
DP[\varnothing] = (0, N, \dots, N),
\]
where $N$ is the total number of tuples in the dataset.
For the state transition, %
let $o_k \notin \mathcal{S}$ be a new physical operator with $impl(o_k) = O_j$.
We define the next operator subset $\mathcal{S}' = \mathcal{S} \cup \{o_k\}$.
Executing $o_k$ incurs a cost proportional to the number of tuples it processes:
\[
C_{\mathcal{S}'} =
C_\mathcal{S} + cost(o_k) \cdot N_j^\mathcal{S}.
\]
The remaining tuples are updated as follows:
\[
N_j^{\mathcal{S}'} = N_j^\mathcal{S} \cdot sel^{intra}_{o_k},\qquad
N_i^{\mathcal{S}'} = N_i^\mathcal{S} \cdot sel^{inter}_{o_k}
\quad \forall i \neq j
\]

For each subset $\mathcal{S}'$, we retain only the transition that minimizes $C_{\mathcal{S}'}$.
Algorithm~\ref{alg:dp-reordering} describes the complete DP procedure.
The DP table is evaluated in increasing order of subset cardinality.
A backpointer table is maintained to reconstruct the optimal operator ordering.
The pseudo-code is below.

\begin{algorithm}[h]
\caption{Dynamic Programming for Operator Reordering}
\label{alg:dp-reordering}
{\footnotesize
\begin{algorithmic}[1]
\Require Physical operators $o = \{o_1,\dots,o_m\}$, logical operators $O = \{O_1,\dots,O_n\}$,
mapping $impl(\cdot)$, costs $cost(\cdot)$, selectivities $sel^{inter}$ and $sel^{intra}$, tuple count $N$
\Ensure Optimal execution order and minimum cost

\State \textbf{Initialize DP table:}
\State $DP[\varnothing] \gets (0, N, \dots, N)$
\State $parent[\varnothing] \gets \bot$

\For{$k = 0$ to $m-1$} 
    \ForAll{$\mathcal{S} \subseteq o$ with $|\mathcal{S}| = k$}
        \ForAll{$o_i \in o \setminus \mathcal{S}$} 
            \State \textbf{Load state from DP table:}
            \State  $(C_\mathcal{S}, N_1^\mathcal{S}, \dots, N_n^\mathcal{S}) \gets DP[\mathcal{S}]$
            \State $\mathcal{S}' \gets \mathcal{S} \cup \{o_i\}$
            \State $O_j \gets impl(o_i)$
            \State $c \gets C_\mathcal{S} + cost(o_i) \cdot N_j^\mathcal{S}$

            \If{$DP[\mathcal{S}']$ is undefined \textbf{or} $c < DP[\mathcal{S}'].cost$}
                \State \textbf{Store updated state in DP table:}
                \State $N_j^\mathcal{S'} \gets N_j^\mathcal{S} \cdot sel^{intra}_{o_i}$
                \ForAll{$\ell \neq j$}
                    \State $N_\ell^\mathcal{S'} \gets N_\ell^\mathcal{S} \cdot sel^{inter}_{o_i}$
                \EndFor
                \State $DP[\mathcal{S'}]  \gets (c, N_1^\mathcal{S'}, \dots, N_n^\mathcal{S'})$
                \State $parent[\mathcal{S}'] \gets o_i$
            \EndIf
        \EndFor
    \EndFor
\EndFor

\State \textbf{Reconstruct optimal order from parent pointers:} 
\State \Return $DP[o]$ and execution order recovered via $parent$
\end{algorithmic}
}
\end{algorithm}

\subsection{Computation of Lower Bounds}
\label{app:lower-bounds}
Here, we present the formulas to compute the lower bounds on precision and recall used in the \emph{Statistical Guarantees} paragraph of Section~\ref{sec:details}.
In the Bayesian framework, we must first formulate our prior beliefs about $\mathrm{Recall}_\mathcal{D}$ with a prior distribution.
As we do not have information about the recall of an arbitrary cascade $P$, we pick an uninformative conjugate $Beta(1, 1)$ prior.
After observing $P_{g}(S)$ on the sample and simulating $P(S)$, we count the true positives and false negatives, which we can use to update our beliefs.
The posterior is given by $\mathrm{Recall}_\mathcal{D} \sim Beta(1 + TP_S, 1+FN_S)$.
The lower bound $\ell_{\alpha_\mathrm{Recall}}$ on recall with credible level $\alpha$ is then given by the equation $\int_{\ell_{\alpha_\mathrm{Recall}}}^1 p(\mathrm{Recall}_\mathcal{D})d\mathrm{Recall}_\mathcal{D}=\alpha$.
Since we use a Beta distribution, we can compute the lower bound using the regularized incomplete beta function:
\begin{align}
\ell_{\alpha_\mathrm{Recall}}
= I^{-1}_{\alpha}\!\left(1+{TP_S},\,1+{FN_S}\right) \label{eqn:recall-lower-bound}
\end{align} 
Similarly, the Bayesian lower bound for precision is given by 
\begin{align}
\ell_{\alpha_\mathrm{Precision}}
= I^{-1}_{\alpha}\!\left(1+{TP_S},\,1+{FP_S}\right) \label{eqn:precision-lower-bound}
\end{align} 
Similarly, for global bounds:
$
\ell_{\alpha_\mathrm{Recall}}
= I^{-1}_{\alpha}\!\left(1+\mathrm{TP}_{global},\,1+\mathrm{FN}_{global}\right)
$ and for precision $
\ell_{\alpha_\mathrm{Precision}}
= I^{-1}_{\alpha}\!\left(1+\mathrm{TP}_{global},\,1+\mathrm{FP}_{global}\right)
$ 

\subsection{Precomputed KV Cache Sizes}
\label{app:kvcache-sizes}
\begin{figure}
  \centering
  \includegraphics[width=1\linewidth]{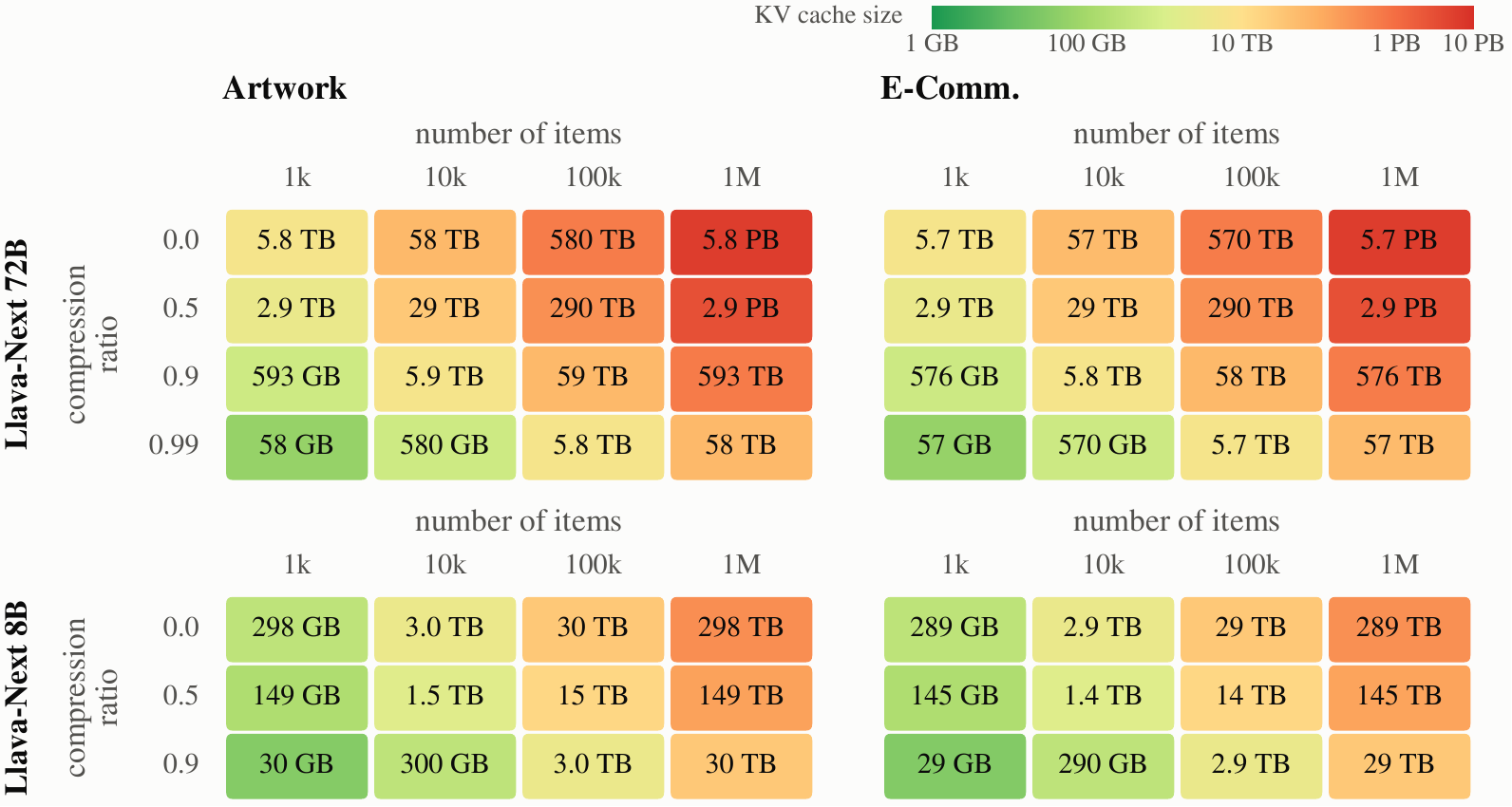}
 \caption{KV cache footprint for visual datasets as a function of item count and compression ratio, color-coded by storage manageability.}
 \label{fig:kv-footprint-overview-image}
\end{figure}

\begin{figure*}
  \centering
  \includegraphics[width=1\linewidth]{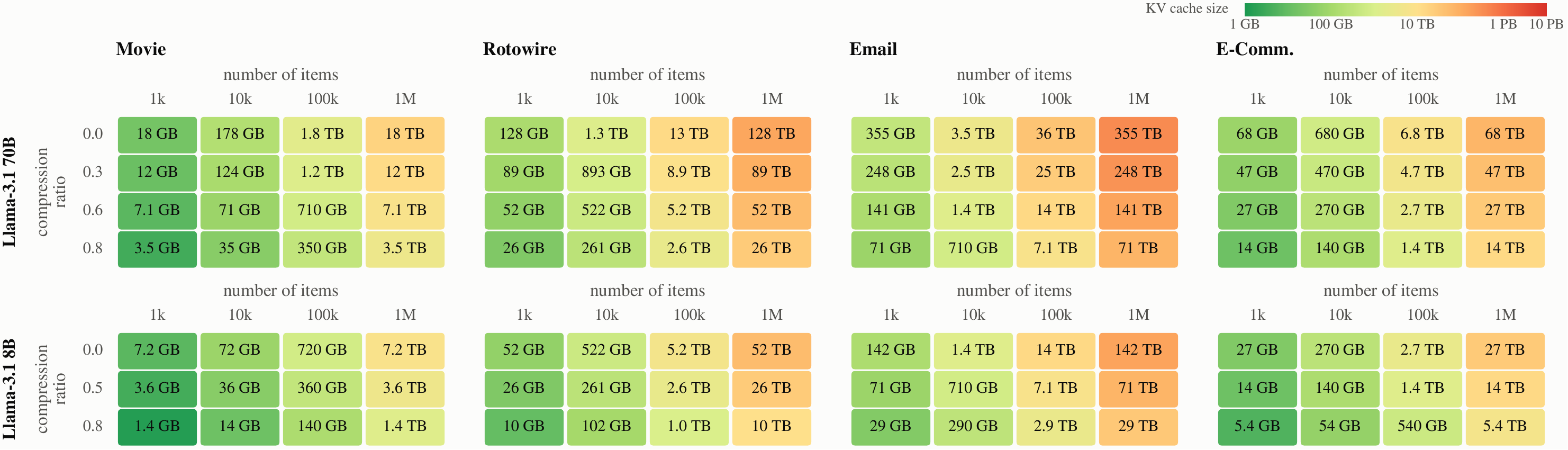}
 \caption{KV cache footprint for text datasets as a function of item count and compression ratio, color-coded by storage manageability.}
 \label{fig:kv-footprint-overview-text}
\end{figure*}

Figures~\ref{fig:kv-footprint-overview-image} and~\ref{fig:kv-footprint-overview-text} illustrate the storage requirements for precomputed KV caches, which are essential for accelerating inference in \ours{}, across textual and visual datasets. We report the storage overhead (in GB) across a range of compression ratios and models, scaling the costs for various dataset sizes. The results are color-coded to highlight manageability, progressing from green for smaller and easily manageable configurations to red for increasingly large caches.

\subsection{Meeting Quality Targets with Increasing Number of Semantic Operators}
\begin{figure}
  \centering
  \includegraphics[width=.48\textwidth]{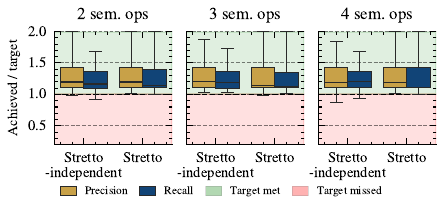}
  \includegraphics[width=.48\textwidth]{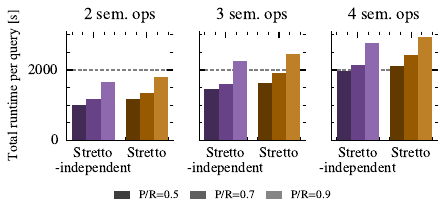}
 \caption{
 Top: \ours{} consistently meets the targets even with a growing number of semantic operators in the query plan. Even with four operators, both the precision and recall target are met. 
 Moreover, we observe that assuming independence is problematic with a higher number of semantic operators, as we stop meeting them at four semantic operators.
 Bottom: runtime grows linearly when increasing the number of operators, as more operators mean more work in both execution and profiling, and because longer pipelines tend to be more selective, leading to more conservative operator picks by the optimizer.
 }
 \label{fig:many-semops}
\end{figure}

Figure \ref{fig:many-semops} shows how runtime and quality change when we add more semantic operators to the semantic query plan.
For this plot, we analyzed our run in Experiment 3 and show that even with a growing number of (logical) semantic operators, \ours{} still meets the targets.
The runtime grows linearly, as more operators means more work for both execution and profiling.
Moreover, since the selectivity can grow as the pipeline contains more filter operators, only few tuples might pass the pipeline, forcing the optimizer to pick operators more conservatively.
Interestingly, if we assume independence of operators, we violate the targets more often when we add more operators, with 7.4\% not meeting the precision target and 6.7\% not meeting the recall target with four semantic operators.
However, assuming independence is only slightly faster than global optimization with more operators, and in most cases not faster at all.

\subsection{Human vs. Golden Plan Labels}
\label{appendix:human-labels}
\begin{figure}
  \centering
  \includegraphics[width=\linewidth]{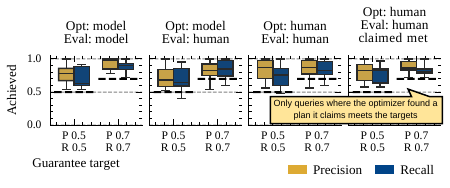}
  \vspace{-7ex}
 \caption{Comparing human vs. golden plan labels. The plot shows different combinations of what labels are used for optimization and evaluation. For example, (Opt: model, Eval: human) optimizes with golden plan labels and evaluates against human labels. This shows that the gold model is a good proxy but \ours{} might miss the target slightly, as human labels for evaluation are stricter.
 When optimizing against human labels, it can happen that targets are genuinely unreachable.
 The difference between (Opt: human, Eval: human) and (Opt: human, Eval: human, \textbf{claimed met}) is that for the latter, we only take into account queries where the optimizer found a plan that guarantees the target.
 In this case, \ours{} guarantees quality w.r.t. the ground truth.
 }
 \label{fig:human-labels}
 \vspace{-2ex}
\end{figure}

Since human labels are typically unavailable for new queries, the optimizer in previous experiments relied on labels of the best available models (i.e., of the golden plan $P_\text{g}$), resulting in guarantees w.r.t. the best model, not the ground truth.
However, in high-stakes domains such as medicine, this might not be sufficient.
In this experiment, we show (1) that optimizing against the golden plan is a good proxy in many cases and that (2) \ours{} can obtain statistical guarantees with respect to the ground truth when the optimizer has access to human labels.
For this we used 15 queries over five datasets for which %
either the dataset provided labels, or we labeled a subset ourselves, as for Artwork.
The results in Figure~\ref{fig:human-labels} show different combinations of which labels are used by the optimizer and which labels are used for evaluation.
When we use the best model labels for optimization, and evaluate on the ground truth, we still meet the target in most cases, but the failures are a bit more frequent, so overall it is a decent proxy.
When we optimize with human labels, it can happen that the target is genuinely not reachable (because the biggest model is not good enough), such that optimizer cannot find a solution.
In that case \ours{} will refuse to execute the query, and the user needs to lower the targets or exchange the best model.
However, if the optimizer finds a solution that it claims meets the targets, as shown in (Opt: human, Eval: human, \textbf{claimed met}) in the Figure, it does meet them with high probability, as before.

